\documentclass[twocolumn]{aastex631}
\pdfoutput=1
\usepackage{comment}
\usepackage{amsmath}	% Advanced maths commands
\usepackage[normalem]{ulem}
\usepackage{amssymb}
\usepackage{natbib}
\usepackage{footnote}
\usepackage{appendix}
\usepackage{microtype}
\usepackage[graphicx]{realboxes}
\usepackage{blindtext}
\usepackage{rotating}
\usepackage{cleveref}
\usepackage{threeparttable}
\usepackage{tablefootnote}

\newcommand\sampsize{9}
\graphicspath{{./}{figures/}}

\newcommand\reff{\ensuremath{r_e}}

\newcommand\physoff{\ensuremath{\delta R}}
\newcommand\reloff{\ensuremath{\physoff/\reff}}

\newcommand\ssfr{\ensuremath{\Sigma_{\rm SFR}}} % Sigma_SFR
\newcommand\msd{\ensuremath{\Sigma_{\rm M*}}} % Sigma_M* 
\newcommand\halpha{\ensuremath{{\rm H\alpha}}}
\newcommand\hbeta{\ensuremath{{\rm H\beta}}}

\newcommand\arm{\ensuremath{{\rm |RM|}}}
\newcommand\rotm{\ensuremath{{\rm RM}}}
\newcommand\rmfrb{\ensuremath{{\rm RM}_{\rm FRB}}}
\newcommand\rmhost{\ensuremath{{\rm RM}_{\rm host}}}
\newcommand\rmcosmic{\ensuremath{{\rm RM}_{\rm cosmic}}}
\newcommand\rmeg{\ensuremath{{\rm RM}_{\rm EG}}}
\newcommand\armeg{\ensuremath{|{\rm RM}_{\rm EG}|}}
\newcommand\rmmw{\ensuremath{{\rm RM}_{\rm MW}}}
\newcommand{\rmunits}{\ensuremath{\rm rad \, m^{-2}}}
\newcommand{\edens}{\ensuremath{n_e}} % Electron density, n_e

\newcommand{\dmunits}{\ensuremath{{\rm pc \, cm^{-3}}}}
\newcommand{\dmcosmic}{\ensuremath{{\rm DM}_{\rm cosmic}}}
\newcommand{\dmacosmic}{\ensuremath{\langle {\rm DM}_{\rm cosmic} \rangle}}
\newcommand{\dmfrb}{\ensuremath{{\rm DM}_{\rm FRB}}}

\newcommand{\dmhost}{\ensuremath{{\rm DM}_{\rm host}}}

\newcommand{\dmhalo}{\ensuremath{{\rm DM}_{\rm halo}}}
\newcommand{\dmism}{\ensuremath{{\rm DM}_{\rm ISM}}}
\newcommand{\dmeg}{\ensuremath{{\rm DM}_{\rm EG}}}

\newcommand\dmhostI{\ensuremath{{\rm DM}_{\rm host,ISM}}}
\newcommand\dmhosta{\ensuremath{{\rm DM}_{\rm host}^{\halpha}}}
\newcommand\dmhostU{\ensuremath{{\rm DM}_{\rm host}^{\rm UV}}}
\newcommand\dmhostC{\ensuremath{{\rm DM}_{\rm host}^{\rm Macquart}}}

\newcommand{\bpar}{\ensuremath{{\rm B}_{\parallel}}}
\newcommand{\bstrength}{\ensuremath{|{\rm B}_{\parallel}|}}
\newcommand{\bunits}{\ensuremath{\rm \mu G}}

\newcommand{\mstar}{\ensuremath{M_*}}  % Stellar mass

\newcommand{\rspear}{\ensuremath{r_S}}
\newcommand{\pspear}{\ensuremath{P_S}}

\shorttitle{FRB Rotation Measure Constraints}
\shortauthors{Mannings et al.}

\begin{document}
\sloppy

\title{Fast Radio Bursts as Probes of Magnetic Fields in Galaxies at $z<0.5$}

\correspondingauthor{A. G. Mannings}
\email{almannin@ucsc.edu}

\newcommand{\UCSC}{\affiliation{Department of Astronomy and Astrophysics, University of California, Santa Cruz, CA 95064, USA}}

\newcommand{\NU}{\affiliation{Center for Interdisciplinary Exploration and Research in Astrophysics (CIERA) and Department of Physics and Astronomy, Northwestern University, Evanston, IL 60208, USA}}

\newcommand{\IPMU}{\affiliation{Kavli Institute for the Physics and Mathematics of the Universe (Kavli IPMU), 5-1-5 Kashiwanoha, Kashiwa, 277-8583, Japan}}

\newcommand{\STSCI}{\affiliation{
Space Telescope Science Institute, Baltimore, MD 21218, USA
}}

\newcommand{\NAOJ}{\affiliation{
Division of Science, National Astronomical Observatory of Japan,
2-21-1 Osawa, Mitaka, Tokyo 181-8588, Japan}}

\newcommand{\JHU}{\affiliation{Department of Physics and Astronomy, Johns Hopkins University, Baltimore, MD 21218, USA}}

\newcommand{\iceland}{\affiliation{Centre for Astrophysics and Cosmology, Science Institute, University of Iceland, Dunhagi 5, 107 Reykjav\'ik, Iceland}
}

\newcommand{\Macquarie}{\affiliation{Department of Physics \& Astronomy, Macquarie University, NSW 2109, Australia}
}

\newcommand{\astrophotonics}{\affiliation{Astronomy, Astrophysics and Astrophotonics Research Centre, Macquarie University, Sydney, NSW 2109, Australia}
}

\newcommand{\CSIRO}{\affiliation{Australia Telescope National Facility, CSIRO Astronomy and Space Science, PO Box 76, Epping, NSW 1710, Australia}
}

\newcommand{\Swinburne}{\affiliation{Centre for Astrophysics and Supercomputing, Swinburne University of Technology, Hawthorn, VIC 3122, Australia}}

\newcommand{\tata}{\affiliation{Department of Astronomy and Astrophysics, Tata Institute of Fundamental Research, Mumbai, 400005, India}}

\newcommand{\NCRA}{\affiliation{National Centre for Radio Astrophysics, Post Bag 3, Ganeshkhind, Pune, 411007, India}}

\newcommand{\PUCV}{\affiliation{Instituto de F\'isica, Pontificia Universidad Cat\'olica de Valpara\'iso, Casilla 4059, Valpara\'iso, Chile}}

\newcommand{\MPI}{\affiliation{Max Planck Institute for Astrophysics, Karl-Schwarzschild-Str. 1, 85741 Garching, Germany}}

\newcommand{\card}{\affiliation{Cardiff Hub for Astrophysics Research and Technology, School of Physics and Astronomy, Cardiff University, Queen’s Buildings, The Parade, Cardiff CF24 3AA, UK}}
\author{Alexandra G. Mannings}
\UCSC

\author{R\"udiger Pakmor}
\MPI

\author[0000-0002-7738-6875]{J.~Xavier~Prochaska}
\UCSC
\IPMU
\NAOJ

\author[0000-0002-6301-638X]{Freeke van de Voort}
\card

\author[0000-0003-3801-1496]{Sunil Simha}
\UCSC

\author[0000-0002-7285-6348]{R.~M.~Shannon}
\Swinburne

\author[0000-0002-1883-4252]{Nicolas Tejos}
\PUCV

\author{Adam Deller}
\Swinburne

\author[0000-0002-9946-4731]{Marc Rafelski}
\STSCI
\JHU

\begin{abstract}
We present a sample of nine Fast Radio Bursts (FRBs) from which 
    we derive magnetic field strengths of the host galaxies
    represented by normal, $z<0.5$ star-forming galaxies with 
    stellar masses $\mstar \approx 10^8 - 10^{10.5} M_\odot$.
    We find no correlation between the FRB rotation measure
    (RM) and redshift which indicates 
    that the RM values 
    %(and derived galactic magnetic field magnitudes) 
    are due mostly to the FRB host contribution. 
    This assertion is further supported by a significant positive correlation
    (Spearman test probability $\pspear <  0.05$) found 
    between RM and the estimated host dispersion measure (\dmhost;
    with Spearman rank correlation coefficient
    $\rspear=+0.75$). 
    For these nine galaxies,  we estimate their magnetic field strengths 
    projected along the sightline $\bstrength$ finding
    a low median value of $0.5 \bunits$.
    This implies the magnetic fields of our sample of hosts are weaker than those characteristic of the Solar neighborhood
    ($\approx 6 \bunits$), but relatively consistent with a lower limit on 
    the observed range of $\approx 2-10 \bunits$ for star-forming, disk galaxies, especially as we consider reversals in the B-field, and that we are only probing \bpar.
    We compare to RMs from simulated galaxies of the Auriga project -- magneto-hydrodynamic cosmological zoom simulations - and find that the simulations predict the observed values to within the $95\%$ CI. 
    Upcoming FRB surveys will provide hundreds of new FRBs 
    with high-precision localizations, rotation measures, and imaging follow-up to support further investigation 
    on the magnetic fields of a diverse population of $z<1$ galaxies.
\end{abstract}


\section{Introduction} \label{sec:intro}
Fast Radio Bursts (FRBs) are milli-second duration pulses of radio emission, arising predominantly from extragalactic sources \citep[e.g.][]{CordesChatterjee19}. 
The first burst discovered, subsequently named the Lorimer burst \citep{Lorimer07}, revealed a new radio transient class with unprecendented power to probe cosmological questions of matter distribution, universal expansion, and (inter-)galactic magnetism due to their dispersion and rotation measures \citep[DMs, RMs; e.g.][]{Gaensler2009, Macquart10, Akahori2011, Macquart2015, Akahori2016}. 
Since the discovery of FRBs, their dispersion measure has already been used to search for answers to long standing questions. Works such as \cite{Macquart20} and \cite{Simha2020} offer 
a nearly complete baryon census---finding baryons where they were once nearly impossible to detect. Galactic halos and the intergalactic medium (IGM) are now being backlit by the flashlights that are FRBs, illuminating the once ``missing" matter. 

FRBs also have potential to probe another influential property of the universe -- magnetic fields \citep[e.g.][]{Piro2018, Hackstein2019}. Many questions such as the origins of magnetic fields, their effects on the evolution of galaxies, and the process of their amplification over cosmological time have been explored extensively
with theoretical treatments \citep[e.g.][]{Springel2010, Pakmor2011, Rodrigues19}. Observational constraints, however,
are currently scant and are critically needed 
to constrain the physical processes at work.
%and therefore test these models. 

There are a number of ways to measure the effects of magnetism in galactic and extragalactic systems, and each method is sensitive to a different magnetic field component (see \cite{Beck2015} for a list of magnetic field components and observational methods). The Zeeman effect can be observed in the emission line spectra of galaxies, indicating a regular field along the line of sight. One can also measure the polarized intensity and linear polarization angle of QSOs and other persistent radio sources, or evaluate signatures from synchrotron radiation which are associated with the magnetic field component that is perpendicular to the line of sight. 

In this study we use the Faraday rotation measures (RMs) ---which quantify the effect of magnetized plasma on linearly polarized radiation--- of FRBs to probe the component of the magnetic field which is parallel to the line of sight (\bpar). 
This measure can elucidate the magneto-ionic environment surrounding the FRB. 
As the signal also interacts with the inter-stellar and circum-galactic media of the FRB host, we can make measurements of the fields in these broader regions as well. 

Constraining these field components helps determine what processes are implicated in the production and amplification of magnetic fields---whether tied to the progenitor object itself and its immediate environment \citep[e.g.][]{Piro2018} or evolution on galactic and cosmic scales \citep[e.g.][]{Hackstein2019}. Comparison of the observed quantities to those predicted by simulations, can be an invaluable test of our understanding of the relationship between magnetic fields and galaxy evolution \citep[e.g.][]{Rodrigues19}. We can also test
FRB progenitor models and how burst properties would be affected. 

In this paper we make use of the Auriga simulations \citep{Grand2017}, a set of high resolution cosmological zoom-in simulations of Milky Way-like galaxies that reproduce many important properties of their observed counterparts. In particular they include a self-consistent model of magnetic field amplification and evolution over cosmic time that produces realistic magnetic field strengths at $z=0$ \citep{Pakmor2017,Pakmor2018,Pakmor2020}. 
We use the simulations to connect the 
FRB observations to conditions
in their local and global environments 
of their host galaxies. 
%We also show that a larger number of bursts can in principle be used to constrain galaxy formation physics.

We also explore the possible connections between the RMs
of a set of FRBs and local characteristics determined by, e.g., \cite{Heintz20}, \cite{Bhandari20a}, and \cite{Mannings21}. 
These works demonstrate that FRBs originate primarily in star-forming
galaxies with stellar masses ranging from 
$\mstar \sim 10^8 - 10^{11} M_\odot$.
These data also reveal the location of the FRBs within their hosts
and constrain local measures such as the star formation density which may correlate
with magnetized plasma.

This paper is organized as follows.
We describe rotation measure in detail in Section~\ref{sec:polar}. In Section~\ref{data}, we outline the selection criteria for our sample (\S \ref{sec:select}), provide a description of rotation measure data (\S\ref{sec:rmdat}), detail the host observations for each burst (\S\ref{sec:host_obs}), and describe host properties (\S\ref{sec:host_prop}).  
In Section~\ref{results}, we discuss the observational analysis and results. We begin by detailing the methods for estimating host contributions to RM and DM 
(\rmhost\ and \dmhost), including the extragalactic contribution to the rotation measure(\S \ref{sec:rm_eg}), the Milky Way contribution to RM (\S \ref{sec:mw_rm}), the correlation between rotation measure 
and redshift 
($\S$ \ref{sec:RMEG_vs_z}), and estimates of \dmhost (\S \ref{sec:dmhost}). We then investigate correlations between \rm and host galaxy characteristics in Section \ref{sec:local_env}, and, in Section~\ref{sec:Bfield}, we make estimates of magnetic field magnitudes of the galaxies hosting the FRBs. 
We then discuss the modeling framework for and results for a magneto-hydrodynamic model of Milky Way-like galaxies (\S\ref{sec:modeling}, \S\ref{sec:auriga_select}) with which we simulate rotation and dispersion measure measurements and compare against observed values (\S\ref{sec:rm_dm}, and \S\ref{sec:rm_dm_auriga}). We finish with a final summary and discussion of implications in Section \ref{sec:discussion}.

\subsection{Polarization and Faraday Effect} \label{sec:polar}

If the oscillations of an electromagnetic field have a preferred orientation, then this radiation is polarized. The polarization of an electromagnetic (EM) wave is determined by the orientation of the electric field component. In general, the polarization of an EM wave is elliptical, i.e. the electric field vector traces an ellipse perpendicular to the propagation direction during transit. Elliptically polarized light can be expressed as a linear combination of two orthogonal linear polarization states or two circular polarization states \citep{griffiths2013}. One requires only three independent parameters to describe the polarization state of an EM wave.

Polarization of radio waves, however, are most often described using the Stokes \textit{I}, \textit{Q}, \textit{U}, and \textit{V} parameters, where \textit{I} refers to the total intensity, \textit{Q} and \textit{U} linear polarization, and \textit{V} circular polarization. These parameters can be combined to represent polarization in the form of the Stokes vector

\begin{equation}
    \vec{S} =
    \begin{pmatrix}
        I\\
        Q\\
        U\\
        V
    \end{pmatrix}
\end{equation}
%with one additional constraint, namely: 

%\begin{equation}
%I^{2} = Q^{2} + U^{2} + V^{2}
%\end{equation} 
%

Estimating \textit{Q}, \textit{U}, and \textit{V} from raw data depends on the specific configuration of the instrument used to detect and measure the polarization signals. As we are including FRBs from multiple experiments across multiple telescopes, descriptions of their methods and parameter formulations can be found in their respective studies \citep{Michilli18, Day20, Mckinven21, Kumar22}. 

The linear polarization angle $\psi$ is expressed as

\begin{equation}
\psi = \frac{1}{2}\arctan\frac{U}{Q} \;\; .
\end{equation}
 $\psi$ is, in general, a function of frequency and time, $\psi(\nu,t)$, and for FRBs it has been observed to evolve over the duration of the burst \citep[e.g.]{Day20, Michilli18}.

As monochromatic light propagates through plasma which has a magnetic field component along the direction of propagation, its linear polarization angle is rotated. The degree of rotation is proportional to the inverse square of the frequency and the proportionality constant depends on the properties of the intervening magnetized medium. While the net rotation at any wavelength or frequency cannot be determined, for a multi-frequency radio signal like an FRB, the rotation measure (\rotm) encodes the properties of the intervening medium and is measured from the variation of the linear polarization angle with wavelength squared:

\begin{equation}
 \rotm = \frac{d\psi}{d\lambda^{2}} \;\;\; .
\end{equation}
 For a pulse of radiation
emitted at redshift \textit{z} that traverses to
Earth, we may express 
%\rotm\ of some object at redshift \textit{z} as 

\begin{equation}
\rotm = C_R \int^0_z \frac{n_e(z) B_\parallel(z)}{(1+z)^2} 
\frac{dl(z)}{dz}dz \, \rm rad \, \rm m^{-2}
\end{equation}

where $C_R$ is a set of physical constants including the inverse square of the electron mass $m_{e}^{-2}$, electron charge cubed $e^3$, and the inverse of the speed of light to the fourth power $c^{-4}$. \edens\ is the electron density, \bpar\ is the magnitude of the line of sight magnetic field, and the integral is over the length of the sightline \textit{dl} with \edens\, \bpar\, and \textit{dl} as functions of z. \rotm\ can be positive or negative depending on the direction of the magnetic field component. In other words, \rotm\ is the average parallel magnetic field strength along the line of sight weighted by \edens.
For FRBs, this includes contributions from the Milky Way, cosmic magnetic fields and the magnetic fields within its host galaxy. 

However, 
it is assumed the field undergoes numerous reversals along the line of sight that minimize the IGM contribution to the RM relative to the host and Milky Way contributions. Our assumptions about the structure of the magnetic field means that our interpretations of RM and derived quantities become model dependent. 
Nonetheless, there exist measurements of RM in cosmic filaments ($RM_f$) such as those presented in \cite{Carretti2022}, which provides estimates around $\rm RM_{f} = 0.71 \pm 0.07 \, \rmunits$. They then infer a magnetic field magnitude in the filaments to be $B_f \approx 32 \rm nG$. 
This value being a tenth of an RM unit (and assuming the value in cosmic voids is even lower due to a lack of ionized material in these regions) motivates an expectation for minimal RM contribution from the IGM. Specific to FRBs, upper limits on the CGM and IGM contributions to FRB Rotation measures, can be found in \cite{Ravi16, Prochaska19RMDM} and \cite{Osullivan20}.

Maps of the Milky Way's magnetic field and Faraday rotation have been developed using  measurements of extragalactic polarized sources, as discussed
in Section~\ref{sec:mw_rm}. 
Once this contribution is subtracted, we can isolate the other components in an effort to better understand magnetic field generation and amplification in the universe, as well as the magneto-ionic environments of FRB progenitors.

\begin{deluxetable*}{cccccccccc}
\tablewidth{0pc}
\tablecaption{FRB Properties and Local Characteristics\label{tab:rm}}
\tabletypesize{\normalsize}
\tablehead{\colhead{FRB} & \colhead{\rmfrb} & \colhead{\dmfrb} & \colhead{$z$} 
& \colhead{\physoff/\reff} & \colhead{\msd} & \colhead{\ssfr}& \colhead{Refs}  
\\& ($\rm rad \, m^{-2}$) & ($\rm pc \, cm^{3}$) & & & ($10^{8} \rm M_{\odot} \, kpc^{-2}$) & ($\rm M_{\odot} \, yr^{-1} \, kpc^{-2}$)
 } 
\startdata 
20121102A$^\dag$&$102700 \pm 100$& 558& 0.193&$0.366 \pm 0.032$&$0.1304 \pm 0.0023$&$4.09 \pm 0.82$&(1),(2)\\ 
20180916B$^\dag$&$-114.60 \pm 0.60$& 349& 0.034&$0.8972 \pm 0.0032$&$0.114992 \pm 0.000021$ &---&(2),(3),(4)\\ 
20180924B&$22.0 \pm 2.0$& 362& 0.321&$1.20 \pm 0.38$&$0.810 \pm 0.010$& $< 0.006$&(2),(5)\\ 
20190102C&$-105.0 \pm 1.0$& 365& 0.291&$0.45 \pm 0.84$&$0.0930 \pm 0.0021$& $< 0.016$&(2),(6)\\ 
20190608B&$353.0 \pm 2.0$& 340& 0.118&$0.88 \pm 0.11$&$0.3398 \pm 0.0011$&$0.0069 \pm 0.0014$&(2),(6)\\ 
20190711A$^\dag$&$9.0 \pm 2.0$& 588& 0.522&$0.6 \pm 1.8$&$0.0448 \pm 0.0038$& $< 0.016$&(2),(6)\\ 
20191001A&$55.50 \pm 0.90$& 508& 0.234&$2.00 \pm 0.14$&$0.3219 \pm 0.0070$& $< 0.005$&(2),(7)\\ 
20200120E$^\dag$&$-29.80 \pm 0.50$& 88& 0.001&$1.607 \pm 0.051$&--- &---&(8)\\ 
20201124A$^\dag$&$-613.0 \pm 2.0$& 411& 0.098&$0.666 \pm 0.013$&--- &---&(9)\\ 
\hline 
\enddata 
\tablecomments{FRB is the TNS name of the fast radio burst.  Those with a dagger are known to repeat. \rmfrb\ is the rotation measure of the FRB.  \dmfrb\ is the dispersion measure of the FRB rounded to the nearest whole number. 
Uncertainties are generally less than 1\dmunits. $z$ is the redshift of the FRB.  
\physoff/\reff\ is the physical offset of the FRB from the host galaxy center in units of effective radii (host-normalized offset). 
\msd\ is the stellar mass surface density of the host galaxy at the FRB location. 
\ssfr\ is the specific star formation rate of the host galaxy at the FRB location. 
FRBs without \ssfr\ or \msd\ values do not have imaging necessary to compute these quantities, and were not reported in \cite{Mannings21}. Refs are the references: (1) \cite{Michilli18} 
(2) \cite{Mannings21} 
(3) \cite{Tendulkar20} 
(4) \cite{CHIME19} 
(5) \cite{Bannister19} 
(6) \cite{Day20} 
(7) \cite{Bhandari20b} 
(8) \cite{Bhardwaj21} 
(9) \cite{Kumar22} 
} 
\end{deluxetable*}

% %%%%%%%%%%%%%%%%%%%%%%%%%%%%%%%%%%%%%%%%%%%%%%%%%%%%%
\section{FRB Data and Sample Selection} \label{data}

\subsection{Selection Criteria} \label{sec:select}

Presently, there are over 600~FRBs in the published literature
and of these $\sim 20$ with published \rotm\ values.
These form the parent sample from which we construct a 
subset for our analysis.
Our scientific foci are to:
\begin{itemize}
    \item Study correlations between local host properties and the inferred host contribution to the RM.
    \item Estimate magnetic fields in FRB hosts.
    \item Make comparisons to cosmological zoom-in simulations that study the relationship between galaxy evolution and magnetic fields.
\end{itemize}
 
These scientific goals helped define the following selection
criteria that each FRB must satisfy:

\begin{enumerate}
    \item A precisely measured \rotm\ value.
    \item A kpc-scale FRB localization precision. 
    \item A high probability association to a host galaxy. 
    \item A spectroscopic redshift measurement for the host galaxy.
    \item Host galaxy imaging and subsequent derived host properties. 
    such as stellar mass, star-formation rate, etc.
    \item Considered and added to the sample by January 2022.
\end{enumerate}

The first criterion is fundamental to the analysis.
The second addresses the fact that
RMs are sensitive to turbulent small-scale magnetic 
fields as well as large-scale ordered fields. 
Requiring kpc-scale localizations allows 
an exploration of correlations between local measures 
such as the star-formation rate surface density 
and \rotm.  In the following analysis, we require 
the net localization uncertainty (statistical and 
systematic error) be less than 5~kpc 
at the redshift of the host galaxy.

Regarding the third criterion, 
we adopt the Probabilistic Association of 
Transients to Hosts  \citep[PATH;][]{Aggarwal_path} 
formalism and demand that the FRB
posterior probability $P(O|x)$ exceeds 95\%.
In general, this criterion is redundant with the 
second as a highly precise localization will generally
yield a secure association provided sufficiently 
deep imaging \citep{Eftekhari20}.
The fifth and sixth criteria allow us to search 
for correlations between the host galaxy properties
and \rotm.

After applying these selection criteria to the
full set of published sources, 
we recover \sampsize~FRBs satisfying the full set.
These are listed in Table~\ref{tab:rm}.

\subsection{Rotation Measures and Other Burst Properties} \label{sec:rmdat}

The \sampsize~FRBs defining our sample are drawn
primarily from two FRB surveys.  The first is 
the Commensal Real-time ASKAP Fast Transients (CRAFT) survey using the Australian Square Kilometre Array Pathfinder (ASKAP) telescope \cite[][]{Macquart10}.  The CRAFT collaboration discovered and observed six of the 
FRB events presented in this paper, on the date in accordance with the TNS name of the event: 
20180924B \citep{Bannister19}, 
20190102C,  20190608B, 20190711A \citep{Day20}, 
20191001A \citep{Bhandari20b}, 
and 20201124A \citep{Kumar22}.  20190711A and 20201124A are repeating bursts whose rotation measure may change with time; the quoted rotation measures are taken from the publications in which these data are presented which are the first detected burst and an average over all detected bursts, respectively. 

Two of the bursts in this sample were detected and characterized by the Canadian Hydrogen Intensity Mapping Experiment (CHIME)/FRB Experiment (FRBs\,20180916B, 20200120E). 
The rotation measure for FRB\,20180916B---located in a nearby spiral galaxy \citep{Tendulkar20}---is presented by the CHIME collaboration with 7 other new (at the time) repeating
FRBs \citep{CHIME19}. The RM for this burst was derived from baseband data collected on FRB\,20181226A, a subsequent repetition of FRB\,20180916B.
FRB 20200120E is localized to a Globular Cluster located in the halo of M81 and is presented in \cite{Bhardwaj21}.

Lastly, we include the source commonly referred to as
``The Repeater'' or R1: FRB\,20121102A.
The rotation measure for FRB\,20121102A was first presented in \cite{Michilli18}, which detailed the extreme magneto-ionic environment in which the burst progenitor must be embedded, in order to produce such a high rotation measure, 
$\rmfrb \sim 10^5 \, \rmunits$. Since this is a repeating burst we take the average quoted in \cite{Michilli18} as our value.

Five out of the \sampsize~FRBs in this sample repeat, leaving four apparently non-repeating bursts. 
The sample has a median $|\rmfrb| = 56 \, \rmunits$
with a range $9 \, \rmunits <  |\rmfrb| < 10^{5} \, \rmunits$
(see Table \ref{tab:rm}).

Repeating bursts can show variability and evolution over the individual burst envelope and with time over subsequent burst repetitions \citep[e.g.][]{Michilli18}. All of the repeating bursts in the sample show at least slight variability in their RMs from burst to burst. 
These variations are insignificant in comparison to the
FRB source to FRB source variation in \rotm\ and do not impact
any of the analysis presented here.

\begin{deluxetable*}{cccccccccccccccc}
\tablewidth{0pc}
\tablecaption{Host Properties\label{tab:hosts}}
\tabletypesize{\normalsize}
\tablehead{\colhead{FRB}  
& \colhead{RA$_{\rm FRB}$} & \colhead{Dec$_{\rm FRB}$} 
& \colhead{RA$_{\rm Host}$} & \colhead{Dec$_{\rm Host}$} 
& \colhead{\mstar} & \colhead{SFR} 
& \colhead{\reff} 
\\&  &  &  &  & ($ 10^{9} M_{\odot}$) & ($M_{\odot} \rm yr^{-1}$) &  (kpc) 
} 
\startdata 
20121102A$^\dag$& 82.9946& $33.1479$& 82.9945& $33.1479$&$0.143 \pm 0.066$ & 0.13&$2.05 \pm 0.11$\\ 
20180916B$^\dag$& 29.5031& $65.7168$& 29.5012& $65.7147$&$2.15 \pm 0.33$ & 0.06&$6.009 \pm 0.012$\\ 
20180924B& 326.1052& $-40.9000$& 326.1052& $-40.9002$&$13.2 \pm 5.1$ & 0.88&$2.82 \pm 0.53$\\ 
20190102C& 322.4157& $-79.4757$& 322.4149& $-79.4757$&$4.7 \pm 5.4$ & 0.86&$5.01 \pm 0.15$\\ 
20190608B& 334.0199& $-7.8982$& 334.0204& $-7.8989$&$11.6 \pm 2.8$ & 0.69&$7.373 \pm 0.059$\\ 
20190711A$^\dag$& 329.4192& $-80.3580$& 329.4192& $-80.3581$&$0.81 \pm 0.29$ & 0.42&$2.48 \pm 0.13$\\ 
20191001A& 323.3513& $-54.7477$& 323.3518& $-54.7485$&$46 \pm 19$ & 8.07&$5.550 \pm 0.029$\\ 
20200120E$^\dag$& 149.4778& $68.8189$& 148.8882& $69.0653$&$72 \pm 17$ & 0.89&$12.50 \pm 0.40$\\ 
20201124A$^\dag$& 77.0146& $26.0607$& 77.0145& $26.0605$&$2.80 \pm 0.50$ & 2.10&$1.988 \pm 0.037$\\ 
\hline 
\enddata 
\tablecomments{FRB is the TNS name of the fast radio burst;  those with a dagger are known to repeat. RA$_{\rm FRB}$,Dec$_{\rm FRB}$ are the coordinates of the FRB. 
RA$_{\rm Host}$,Dec$_{\rm Host}$ are the coordinates of the host galaxy. 
\mstar\ is the stellar mass of the host galaxy.  
SFR is the star formation rate of the host galaxy, with typical uncertainty of 30\%\ (systematic). 
\reff\ is the effective radius of the host galaxy.  
} 
\end{deluxetable*}

% %%%%%%%%%%%%%%%%%%%%%%%%%%%%%%%%%%%%%%%%%%%%%%%%%%%%%%%%%%%%%
\subsection{Host Observations}
\label{sec:host_obs}

Nearly all of the observations of the host galaxies
for our RM sample have been published previously.
Here we briefly review the primary datasets.

Regarding imaging, where available we have leveraged
high-spatial resolution data obtained with the
Hubble Space Telescope (\textit{HST}).
Six of the hosts in the sample were observed by 
{\it HST} and its 
Wide-field Camera 3 (WFC3) in
UVIS and IR images (F300X and F160W filters, respectively) taken as part of GO programs 15878 (PI: Prochaska) and 16080
(PI: Mannings).

These programs targeted galaxies for which FRB events have been detected and localized by the CRAFT survey. 
These images were previously published in \cite{Chittidi21} and \cite{Mannings21}. 
Information for FRBs~20180924B, 20190102C, 20190608B, 20190711A, and 20191001A was drawn from this dataset. 

We also include {\it HST} images from GO program 14890 (PI: Tendulkar) which observed the host of FRB~20121102A \citep{Bassa17}. 
These observations include images taken in the F110W and F160W IR filters (equivalent to {\it J} and {\it H} bands, respectively), and a narrow-band H-$\alpha$ image with the F763M filter. Detailed descriptions of image processing and reduction can be found in \cite{Bassa17} and \cite{Mannings21}. 

The high spatial-resolution imaging is complemented by 
multi-band, ground-based images from public surveys
and directed follow-up campaigns.
We refer the reader to 
\cite{Heintz20} and \cite{ Bhandari20a, Bhandari20b, Bhandari22} for details and note
the data are all taken from the FRB repository
on GitHub \citep{frb_repo}.

\begin{figure}[ht]
    \centering
    \includegraphics[width=\linewidth]{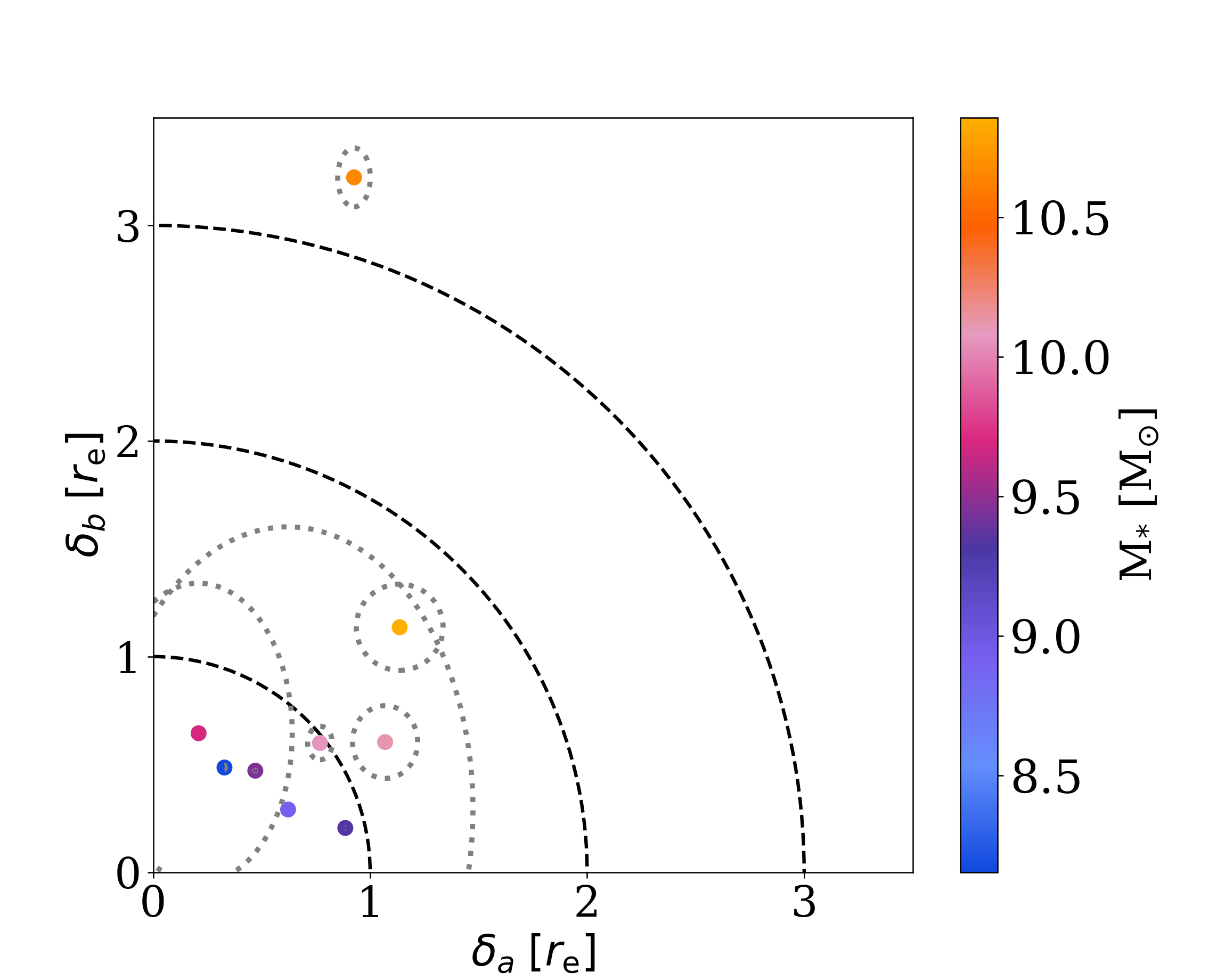}
    \caption{
    Locations of the FRB sample relative to their host
    galaxy centroids along the 
    major ($\delta_a$) and minor ($\delta_b$) 
    axes in units of \reff, defined by the half-light radius determined by {\sc galfit}. 
    The gray dotted ellipses around each of the points show the FRB localization error relative to the size of the host (several are smaller than the symbols). 
    In the case of FRB\,20190711A, the FRB localization is almost 2 times the effective radius of the host, resulting in a fairly large ellipse around the central point. FRB\,20191001A is highly offset along the minor axis ($\delta b \approx 3$), but less-so along the major axis ($\delta a \approx 1$). See images in \cite{Mannings21}.
    The points %are colored by their host stellar mass and 
    cluster at $\delta_a, \delta_b < 1.5 \reff$. Therefore, the bursts are predominantly within 
    the inner disks of the galaxies but rarely 
    (if ever) from the nucleus 
    itself  (i.e. $<<r_e$). 
    }
    
    \label{fig:location}
\end{figure}

\subsection{Host Properties} \label{sec:host_prop}

Central to our study is an exploration of the properties
of the galaxies hosting the FRBs, both global and local
measures.   
We use the quantities derived from previous studies 
throughout this work:
star-formation rates
(global and local to the FRB,
SFR and \ssfr), 
effective radii (\reff), 
stellar mass (global \mstar\ and 
local \msd\ to the FRB), 
and offsets.
These are tabulated in Tables~\ref{tab:rm} (local properties) and
\ref{tab:hosts} (global galaxy properties).

Figure~\ref{fig:location} shows the locations of the FRBs
within their host galaxies relative to the host galaxy
centroid and in units of \reff.
We have de-projected the offsets
along the major and minor axes using fits to each host 
with the \texttt{galfit} \citep{GALFIT} software package.
Most of which are reported in \cite{Mannings21}, with the remaining two fits
being presented here (FRBs\,20200120E and 20201124A).

We observe that 
most of the bursts are located
within $\approx 1.5 \reff$ from the centers of their hosts, 
with one burst residing further out in its host's disk at $\approx 3 \reff$. 
Furthermore, \cite{Mannings21} characterizes the FRB locations in that sample as occurring at moderate offsets on or near spiral arm structure. \cite{Bhardwaj21} shows that FRB\,20201001E likely originated in a globular cluster in the outskirts of M81. In contrast, FRB\,20121102A occurs very near a central star-forming region in its host.
%In \citep{Mannings21} the authors show that
Therefore, as regards the ISM contribution to the \rotm,
one expects variation as the bursts occur in relatively diverse environments although the majority are located on or near spiral structure.
%in rotation measures due to the immediate environments around the progenitors and the broader differences and variations across different regions of the ISM.
%but some variation  
%See Section \ref{sec:Bfield} for a more detailed discussion of the implications of this observation. 

\begin{figure}[ht!]
        \centering
        \includegraphics[width=0.5\textwidth]{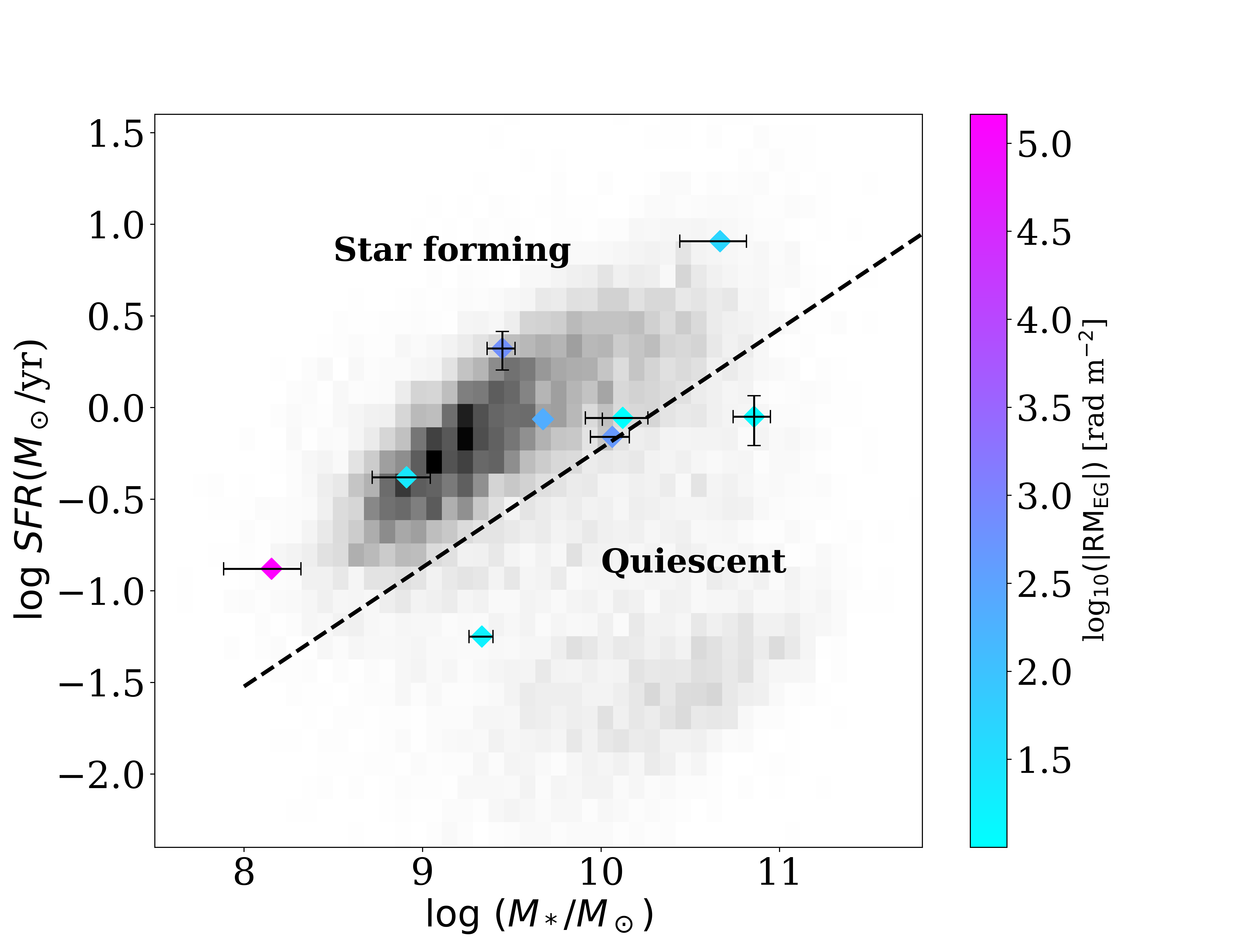}
        \includegraphics[width=0.5\textwidth]{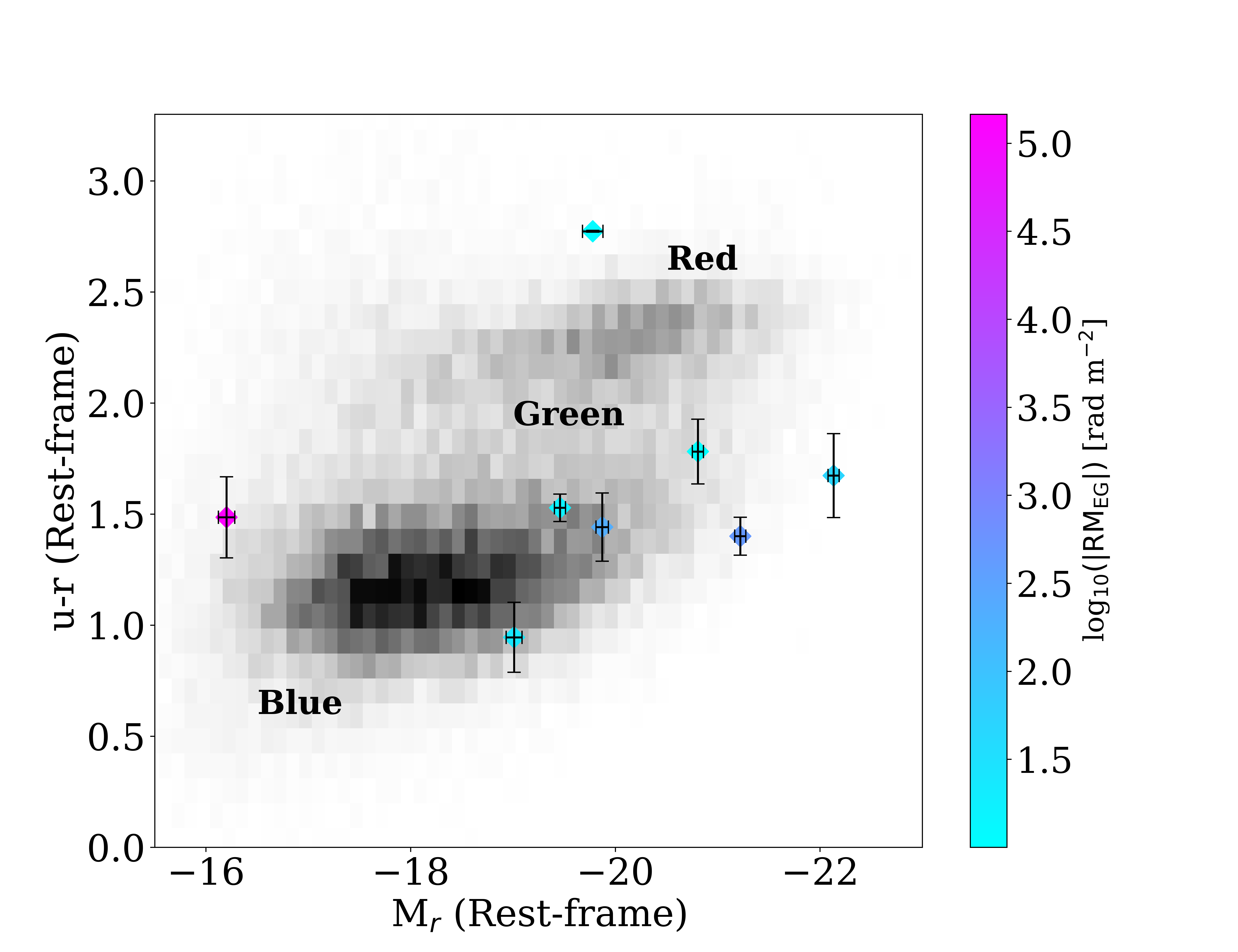}
        \caption{
        (top)
        Diamond points (colored by  $\log_{10} |\rmeg|$; the RM component excluding Milky Way contribution; see $\S$ \ref{sec:mw_rm})
        present the global measures of stellar
        mass (\mstar) and star-formation rate (SFR)
        for the host galaxies of our RM sample.
        The dashed line divides quiescent and star-forming hosts, and the 2D histogram
        describes the distribution for field galaxies from the PRIMUS survey \citep{Primus}. 
        The majority of hosts lie in the 
        star-forming region of the diagram with two in 
        the quiescent region.
        (bottom)
        Color-magnitude diagram with points and histogram as above. 
        The labels ``Blue", ``Green", and ``Red" 
        specify in turn:
        blue galaxies thought to be currently star-forming with younger stellar populations, 
        green hosts transitioning out of star-formation, and ``red and dead" galaxies where 
        star-formation has ceased and older stellar populations dominate the host. 
        The majority of our sample populate the blue cloud or green valley indicative
        of moderate star-formation.
        }
        \label{fig:global_host}
\end{figure}

We also characterize the host galaxies in the sample according to their overall properties. In Figure~\ref{fig:global_host}, we compare the 
global properties of the host galaxies in the
sample against measurements of field galaxies at 
similar redshift. In the bottom panel, we show a 
color-magnitude diagram where hosts in the ``Blue" region are early-type galaxies with 
young stellar populations and active star-formation, 
while those in the ``Red" region are late-type hosts 
with very low star-formation and older stellar populations. 
The majority of these FRB hosts
are star-forming and lie either in the so-called 
blue-cloud of galaxies or the green valley, as supported by what is shown in the upper panel (the star-formation rate vs stellar mass diagram)
where most of the hosts reside in the "star-forming" region of the plot.
The two notable exceptions are 
FRB~20200120E and FRB~20180916B which have hosts with
non-zero SFR but lie below the  
SFR main-sequence,
%In the top panel of Figure~\ref{fig:global_host} (the star-formation rate - stellar mass diagram), we see that the most of the hosts are in the ``star forming" region of the plot, with two in the ``quiescent region" and two transitioning.
The figure indicates that the galaxies studied
here have properties typical of a $z \sim 0.2$
population with a preference for star-forming
and more luminous/massive galaxies.
For the full parent population of FRB host
galaxies, however, \cite{Safarzadeh20} 
demonstrated that the hosts are less massive
(and have lower SFR) than a sample weighted
by SFR. The points are colored by the extra-galactic RM (\rmeg; see eq.\ref{eqn:rmeg}), but there is no apparent correlation between the host properties and \rmeg.

% %%%%%%%%%%%%%%%%%%%%%%%%%%%%%%%%%%%%%%%%%%%%%%%%%%%%%%

\section{Observational Analysis and Results} \label{results}

In this section we analyze the \rotm\ measurements
to search for correlations with the host galaxy
properties and to estimate the underlying magnetic
fields within the hosts. We  begin by
introducing approaches to isolate the RM and
DM contributions from the host.

The host and burst characteristics for FRB\,20121102A are anomalous in comparison to the rest of the sample, as the host is a star-forming dwarf with a persistent radio source, and the burst's \rmfrb\ 
is orders of magnitude higher than other bursts in this sample. 
Much attention has been given to FRB\,20121102A with respect to its high \rmfrb\ , in an attempt to determine what connection this value has to possible progenitor channels and local magnetic field properties. 

 It should be noted that the \rmfrb\ of 20121102A is not completely unique with the detection of FRB\,20190520B \citep{Zhao_2021, Niu2022} whose \dmfrb\ and \rmfrb\ are both much higher than what has been observed with other bursts. The \rmfrb\ varies substantially, but reaches a maximum $ > 1.3 \times 10^4\ \rmunits$ \citep{Annathomas23}, and the progenitor argued to be embedded in a combined magnetar wind nebula and supernova remnant \citep{Zhao_2021}.
We include FRB\,20121102A in our analysis despite its extreme RM,
but, where relevant, we comment on results without its inclusion.

\subsection{Constraining the Contribution to \rmfrb\ from the IGM} \label{sec:rm_eg}

In this subsection, we search for any trend of \rmeg with redshift
akin to the Macquart Relation for the dispersion measure. 

In this case we do not expect a trend with redshift since there is no expected preferred direction of magnetic fields on cosmic scales. The random orientation of these intergalactic fields, and probable dominance of the turbulent field components, leads to field reversals along what can be considered to be a random walk--- where the mean field is $ \langle B \rangle \sim 0$, while $\sigma_{B} ^{2} \ggg 0$. Therefore, integrated along the line of sight, 
\bpar\ approaches zero on average (\S \ref{sec:polar}).
We first, however, describe our approach to removing an estimated
contribution to \rmfrb\ from our Galaxy.

\subsubsection{Milky Way Rotation Measure (\rmmw)} \label{sec:mw_rm}

{Each of the \rmfrb\ measurements include a 
contribution from the path through our galaxy.
This includes both the interstellar medium
and any halo component.

The Faraday map presented in \cite{Oppermann12} uses surveys of polarized extragalactic radio sources to determine rotation measures within and outside of the Galactic plane.

This model's methodology and theoretical framework is used as a basis for the production and improvement of the HE20 model \citep{Hutsch20, Hutsch21}---a Faraday sky model that uses the correlation between Galactic Faraday rotation and Galactic free-free emission to increase the accuracy of previously developed maps \citep{Hutsch20}. They also incorporate a new all-sky data set, which has a higher density of sources near the galactic plane and other under-resolved areas of the sky (such as the southern sky). These improvements increase the resolution of the resulting maps by a factor of two over previous studies \citep{Hutsch21}.

Therefore, we use the HE20 model of \rmmw\ to account for 
the Milky Way's contribution to \rmfrb\ 
to thereby isolate the extragalactic RM contribution:

\begin{equation}
\rmeg = \rmfrb - \rmmw
\label{eqn:rmeg}
\end{equation}
Last,  we note that maps such as these are limited in their spatial resolution in regards to particular lines of sight through the Milky Way. However, the HE20 model uses a total of 55,190 sources (primarily from the LOFAR Two-metre Sky Survey and NRAO VLA Sky Survey RM catalogs), resulting in improvements in resolution and uncertainties.
%resolution of $\simeq 46.8 \rm arcmin^{2}$. 
The uncertainties mostly range from $80 \rmunits$~in the plane of the galaxy to
$\sim ~0 \rmunits$~as we move to lines of sight further from the disk. 
There are few regions in the Galactic plane (specifically towards Galactic center) where the uncertainties reach$~\simeq 300$ \rmunits \citep{Hutsch21}, but none of our FRBs have LOS near $|l| \sim 0^\circ$ .

\begin{figure}[h!]
    \centering
    \includegraphics[width=\linewidth]{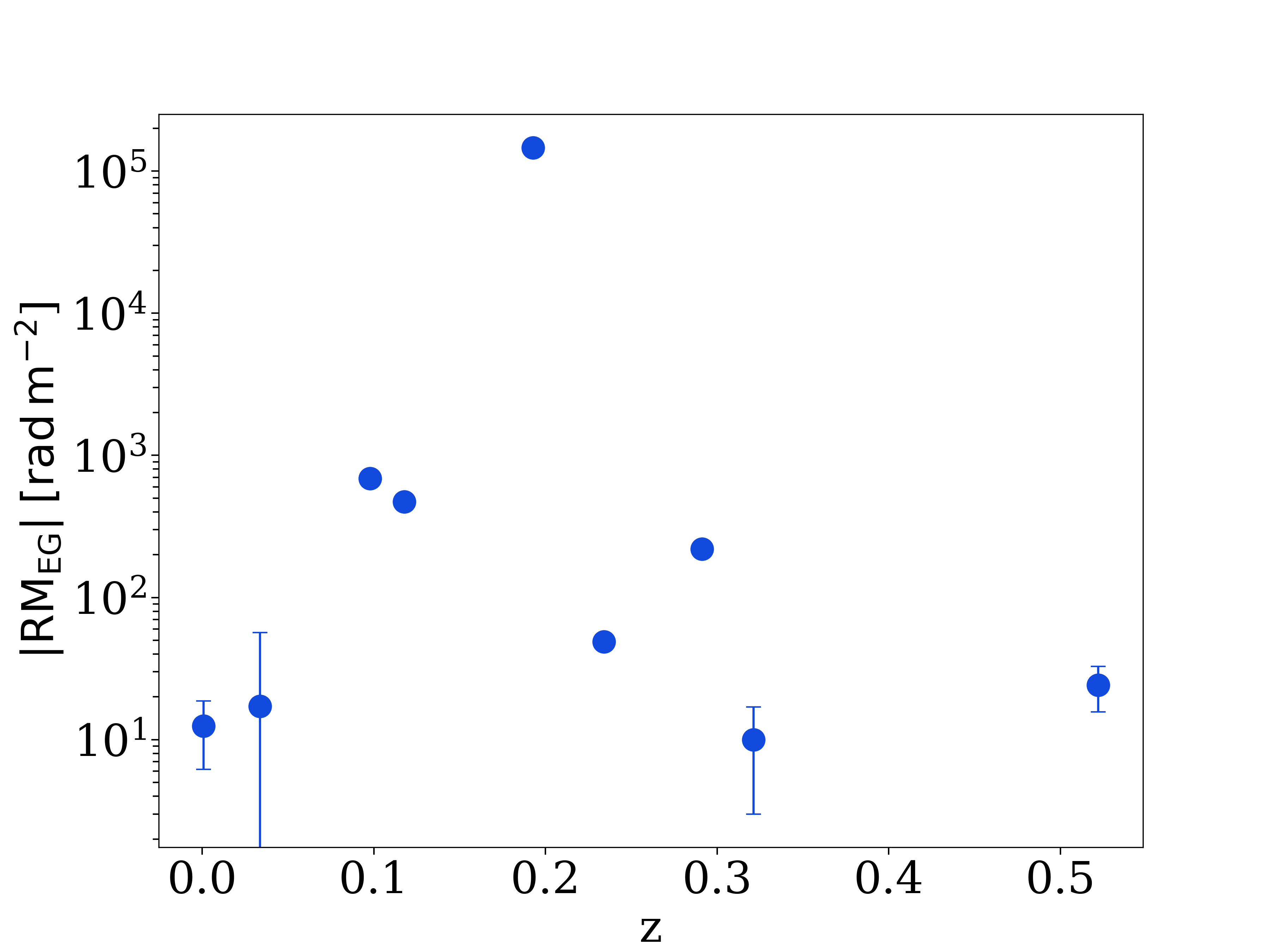}
    \caption{
    Amplitude of the extragalactic
    rotation measure (\armeg) 
    as a function of redshift ($z$) for the full set of 
    FRBs in this sample. Error bars are shown, where some errors are smaller than the points.
    There is no significant correlation between the two,
    in sharp contrast to the burst DM which shows a clear dependence on redshift 
    \citep[evidenced in the Macquart Relation;][]{Macquart20, James22a}. 
    This indicates no strong influence from 
    intergalactic magnetic fields and that 
    the local environment (host and immediate burst environment) 
    dominates \armeg.
    The extremum of the \armeg\ distribution is
    FRB\,20121102A which is expected to reflect a highly 
    magnetized environment in which the burst resides.
    }
    \label{fig:RM_vs_z}
\end{figure}

\subsubsection{Correlating \rmeg\ with $z$}
\label{sec:RMEG_vs_z}

Figure~\ref{fig:RM_vs_z} plots the absolute value
of the extragalactic \rmeg\ for each sightline
against the FRB redshift.
There is no discernible trend between these two
quantities.  Both parameter (slope of the best-fit
line) and non-parametric (Spearman) tests reveal
no significant correlation.  
This stands in stark contrast to the strong correlation
observed between DM and redshift (the Macquart relation), which  arises
from the highly ionized cosmic web
\citep{Macquart20}. 
The absence of any apparent correlation 
in Figure~\ref{fig:RM_vs_z} implies the 
IGM makes little contribution to the overall RM,
primarily due to reversals in the magnetic field over cosmic scales.
This
is consistent with upper limits for the
contribution of intervening galaxy halos 
\citep{Lan2020}.

For the remainder of the paper, we assert that 
the cosmic contribution to \rmfrb\ 
(\rmcosmic) is negligible.  Therefore
\rmfrb\ is dominated by only two components along
the sightline, our Galaxy and the FRB host.
In turn, this implies
%$\rmeg \approx \rmhost$, 
\rmeg\ is dominated by the host contribution
and given  equation~\ref{eqn:rmeg} we have,

\begin{equation}
\rmhost \approx (\rmfrb - \rmmw) \, (1+z)^2  \;\;\; ,
\label{eqn:rmhost}
\end{equation}
which explicitly applies a factor 
of $(1+z)^2$ to correct to the host rest-frame. 
In what follows,
we test equation~\ref{eqn:rmhost}
and then derive estimates for the 
magnetic field strength of the FRB host galaxies
based on its evaluation.

% %%%%%%%%%%%%%%%%%%%%%%%%%%%%%%%%%%%%%%%%%%%%%%%%%%%%%%%%%%%%
\subsection{Estimating the Host Dispersion Measure (\dmhost)}
\label{sec:dmhost}

\begin{deluxetable*}{cccccccc}
\tablewidth{0pc}
\tablecaption{Estimated RM, DM and Magnetic Field of the Host Galaxies\label{tab:B}}
\tabletypesize{\normalsize}
\tablehead{\colhead{FRB} & \colhead{\rmhost}
& \colhead{\rmmw} 
& \colhead{\dmhosta} & \colhead{\dmhostU} 
& \colhead{\dmhostC} 
& \colhead{\bstrength} 
\\& (\rmunits) & (\rmunits) 
 & (\dmunits) & (\dmunits) 
 & (\dmunits) & (\bunits)
 } 
\startdata 
20121102A$^\dag$&$146127 \pm 137$&$-18 \pm 37$& 183& 3519& $231 \pm 99$&$777 \pm 366$\\ 
20180916B$^\dag$&$-17 \pm 40$&$-99 \pm 39$& 167 && $113 \pm 48$&$0.19 \pm 0.43$\\ 
20180924B&$10.0 \pm 7.0$&$16.3 \pm 5.0$& 311& $<117$& $-3 \pm 205$&$0.4 \pm 2.8$\\ 
20190102C&$-219.5 \pm 8.7$&$26.6 \pm 7.7$& 177& $<202$& $17 \pm 163$&$9 \pm 54$\\ 
20190608B&$472 \pm 15$&$-24 \pm 13$& 135& 154& $181 \pm 28$&$3.2 \pm 1.1$\\ 
20190711A$^\dag$&$-24.2 \pm 8.5$&$19.4 \pm 6.5$ && $<173$& $34 \pm 333$&$0.9 \pm 9.2$\\ 
20191001A&$48.8 \pm 5.2$&$23.5 \pm 4.3$& 512& $<116$& $272 \pm 126$&$0.22 \pm 0.11$\\ 
20200120E$^\dag$&$-12.4 \pm 6.3$&$-17.4 \pm 5.8$ & && $7 \pm 4$&$0.51 \pm 0.26$\\ 
20201124A$^\dag$&$-686 \pm 26$&$-44 \pm 24$& 810 && $181 \pm 2$&$4.7 \pm 1.2$\\ 
\hline 
\enddata 
\tablecomments{Daggers denote repeating FRBs. \dmhostC for FRBs\,20180924B, 20190102C, and 20200120E are all below 30.For the calculation of \bpar we set a minumum DM of 30 \dmunits, but report the derived values here. Those left blank do not have the necessary measurements to calculate \dmhost using the particular method.} 
\end{deluxetable*}

In section~\ref{sec:RMEG_vs_z}, we argued that the extragalactic
RM (\rmeg) estimates for the FRBs arise from ionized 
and magnetized gas within their
host galaxies. 
%(i.e., ${\rm RM}_{\rm EG} \approx \rmhost$).  
Adopting this expectation, we may leverage
observations of the galaxies and the local environments 
of the FRBs to provide further insight into the 
underlying magnetic field. 
In particular, we aim to 
provide an order-of-magnitude estimate for the magnetic field
strength.  Following standard treatment for sightlines through the
Galactic ISM \citep[e.g.][]{Arshakian2009, Beck2019}, one requires an estimate
for the dispersion measure of the gas giving rise to \rmhost to calculate \bstrength.
We will also use this DM estimate to guide the
models of \rmhost\ presented in Section~\ref{sec:modeling}.

We will make the further assumption that \rmhost\ is dominated 
by gas within the host galaxy ISM and/or the local environment
of the FRB.  Specifically, we ignore any 
RM contribution from 
the diffuse and ionized gas of the host halo.  This is due to the
lower anticipated density of halo gas, as supported by galaxy
formation theory and simulations (see $\S$~\ref{sec:modeling}).
Therefore, we wish to estimate the dispersion measure of the
host foreground to the FRB.
%which we refer to as \dmhost.
We define this quantity as \dmhostI\ and assume that $\dmhost \approx \dmhostI$.

We consider several approaches to estimate \dmhostI, each of which
bears significant uncertainty.
One approach follows \cite{Reynolds1977} (further developed in \cite{Tendulkar17}) who introduced a method
to relate the emission measure EM ($\propto n_e^2$) to
the sightline DM ($\propto n_e$) allowing for a parameterization
of the unknown clumping of the gas. 
For EM, we consider two observed fluxes of radiation from
gas towards the FRB.

The first EM is the observed surface
brightness of Hydrogen recombination radiation (e.g.\ \halpha).
In its favor,
the majority of FRB host galaxies have one or more optical,
nebular Hydrogen recombination lines measured from optical 
spectroscopy \citep[e.g.][]{Bhandari22}.
On the negative side, most of these were obtained from 
long-slit observations centered on the host galaxy and 
not necessarily including the FRB location.  Furthermore,
the typical atmospheric seeing of 
$\sim 1''$ and the generally small
angular sizes of the galaxies 
(with exceptions)
yield only a characteristic surface brightness from the
host galaxy ISM.  
Table~\ref{tab:B} lists a set of estimates of \dmhosta\ 
based on published, integrated \halpha\ flux 
measurements\footnote{Or \hbeta\ converted to \halpha\ 
using standard nebular flux ratios.}, 
the angular sizes of the galaxies, and 
corrected for dust extinction.
These range from one to many hundreds \dmunits.

For the subset of FRBs with hosts observed at high spatial
resolution, we may better constrain the emission measure
at the FRB location.
Six of the \sampsize\ FRBs have extant Hubble Space Telescope
(HST) observations at $\approx 0.1''$~FWHM resolution
\citep{Mannings21}.  
These are primarily
broadband images at UV and near-IR wavelengths. 
The UV emission is dominated by radiation from massive
stars which also drive the nebular Hydrogen emission and
the two are strongly correlated \citep[e.g.][]{calzetti01}. 
For those with UV, we
use the standard scaling between near-UV luminosity and
\halpha\ luminosity \citep{Kennicutt98} to 
estimate the \halpha\ surface brightness from the UV observations.
We then calculate the emission measure and relate this to a dispersion measure estimate.
% {\bf We apply a dust correction to the H-$\alpha$ surface brightness using the total extinction ($A(V)$) --- including this correction in the calculation of the emission measure.}
%[what do we do for dust extinction?  Check what we did in Chittidi]
Table~\ref{tab:B} lists the \dmhostU\ estimates for these
six galaxies.

\begin{figure*}[!ht]
    \centering
    \includegraphics[width=0.8\textwidth]{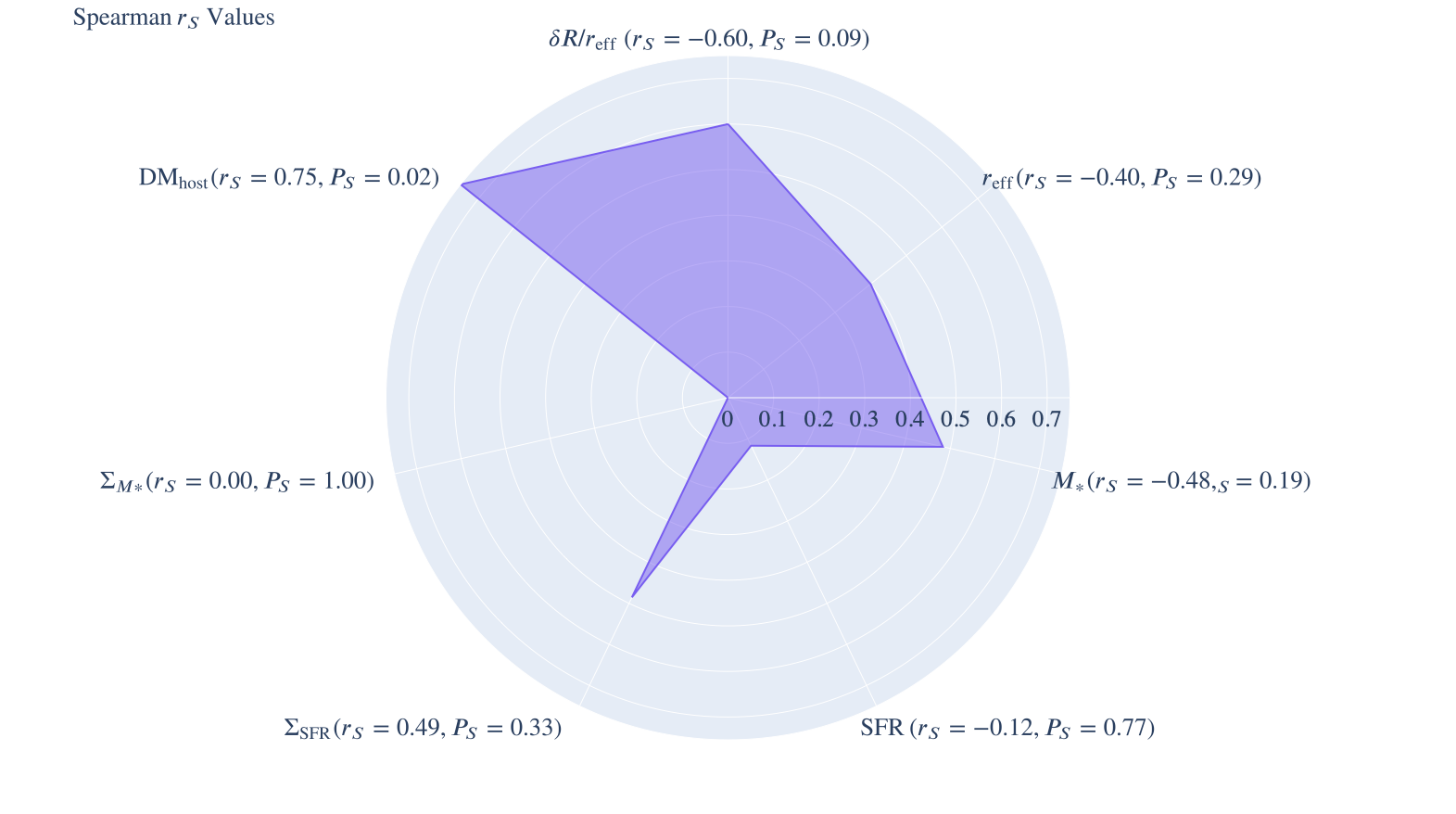}
    \caption{A radar plot showing Spearman \rspear\ values as tests for the correlation between 
    absolute rotation measure (\arm) and a range
    of galaxy-scale and local environment 
    related measures. The radial values represent show the absolute value of the stated \rspear\ values, and the associated significance measures are shown as well. A negative \rspear\ represents anti-correlation.
    The only correlation meeting our significance criterion ($\pspear < 0.05$) is that with \dmhost.
    This positive correlation can naturally arise from the
    fact that both quantities are sensitive to
    the electron density in the host ISM. 
    %and it's contribution to these
    %integrated quantities (\rmhost and \dmhost). 
    }
    \label{fig:radar}
\end{figure*}

Last, we consider a complementary approach to estimating \dmhostI\ 
using the Macquart Relation which relates the cosmic dispersion
measure \dmcosmic\ with redshift \citep{Macquart20}.
We refer to this estimation as \dmhostC.
The method is to subtract from the observed total dispersion measure
\dmfrb\ estimates for the Galactic ISM and halo
(\dmism, \dmhalo) and the cosmic web (\dmcosmic):

\begin{equation}
\dmhostC = \dmfrb - \dmism - \dmhalo - \dmcosmic
\label{eqn:host}
\end{equation}
We use the NE2001 model \citep{ne2001} of the Galactic ISM
to evaluate \dmism\ for each FRB sightline 
and assume the Milky Way halo
contributes $\dmhalo = 40 \dmunits$ \citep[e.g.][]{xyz19}.
From the host galaxy redshift, we calculate
the average \dmacosmic\ \citep{Macquart20}.
This yields the \dmhostC\ values listed in Table~\ref{tab:B} which have also been corrected
to the rest frame of the host (i.e.\ we applied
a factor of 1+$z$).
Formally, these include gas from both the ISM and halo
of the host galaxy and current work suggests the host halo
term may contribute several tens \dmunits\ \citep{xyz19}.
We also note that structure in the cosmic web
lends to an asymmetric scatter about \dmacosmic\ 
and the median \dmcosmic\ value is predicted to be several
tens \dmunits\ lower than the mean.
These two corrections offset against one another, 
and we therefore
proceed with equation~\ref{eqn:host}
for our \dmhostC\ estimates acknowledging that these
bear $\sim 30 \dmunits$ uncertainty.

Inspecting the values of \dmhostI\ listed
in Table~\ref{tab:B}, one notes each approach
exhibits a distinct distribution.  The
\dmhosta\ estimations span the largest range
and exhibit the largest values.  
Indeed, many exceed \dmhostC\ and even the total
\dmfrb\ of the sightline.  The dust-corrected
\dmhostU\ distribution are primarily upper limits
of one hundred to a few hundreds \dmunits\ and are generally
consistent with \dmhostC.
We proceed with analysis that adopts \dmhostC\ 
for the host dispersion measure, however of the values calculated 
for \dmhostC are $\sim 1$ or negative. Therefore 
we impose a minimum value of 30~\dmunits\ 
based on the minimum value of our own Galactic
ISM (NE2001), noting that all of the galaxies
exhibit signatures of star-formation and must
harbor a non-negligible ISM.
This minimum value also accounts for uncertainty
in the methodology. For completeness, the values presented in Table~\ref{tab:B} 
show the calculated values for \dmhostC including those which are below this minimum. However, the minimum
threshold is implemented in all subsequent analysis.

\subsection{Correlating Host Characteristics 
with Rotation Measure}\label{sec:local_env}

We now test for correlations between global and local characteristics of the host galaxies and RM isolated
to the host galaxy (i.e.\ with the Galactic component
subtracted using equation~\ref{eqn:rmhost}; \rmhost). 
To the extent that \rmhost\ traces gas beyond the local
environment, we may identify correlations with the
host galaxy properties.
For example, one may expect FRBs found in regions of 
elevated star-formation to exhibit a higher rotation measure. Magnetic fields get wrapped up into forming stars, but the fields can also be amplified by ionizing radiation and turbulence from violent star formation, cloud collapse, and massive star death.

Specifically, we consider global
measures of the star-formation rate (SFR), 
stellar mass (\mstar), and
galacto-centric offset (relative to the galaxy
effective radius, \reloff). 
We also consider local measures of the 
SFR and \mstar\ surface densities 
(\ssfr, \msd; Table~\ref{tab:rm})
and our estimate of \dmhost\ 
which would include both local and ISM contributions.

We perform Spearman tests - computing Spearman 
correlation coefficients \rspear\ - with the null hypothesis that there is no correlation between rotation measure and a given measurement. 
We select the Spearman test because there is no assumption of Gaussian distributions for the variables. 
We perform the analysis in log-log space (as opposed to linear space) 
as the power-law relationships appear linear, and 
many of these values span several orders of magnitude. 
Last, we set a threshold of the Spearman probability
for a significant correlation at $\pspear < 0.05$ 
(i.e.\ requiring 95\% significance).

\begin{figure*}[!ht]
    \centering
    \includegraphics[width=0.9\textwidth]{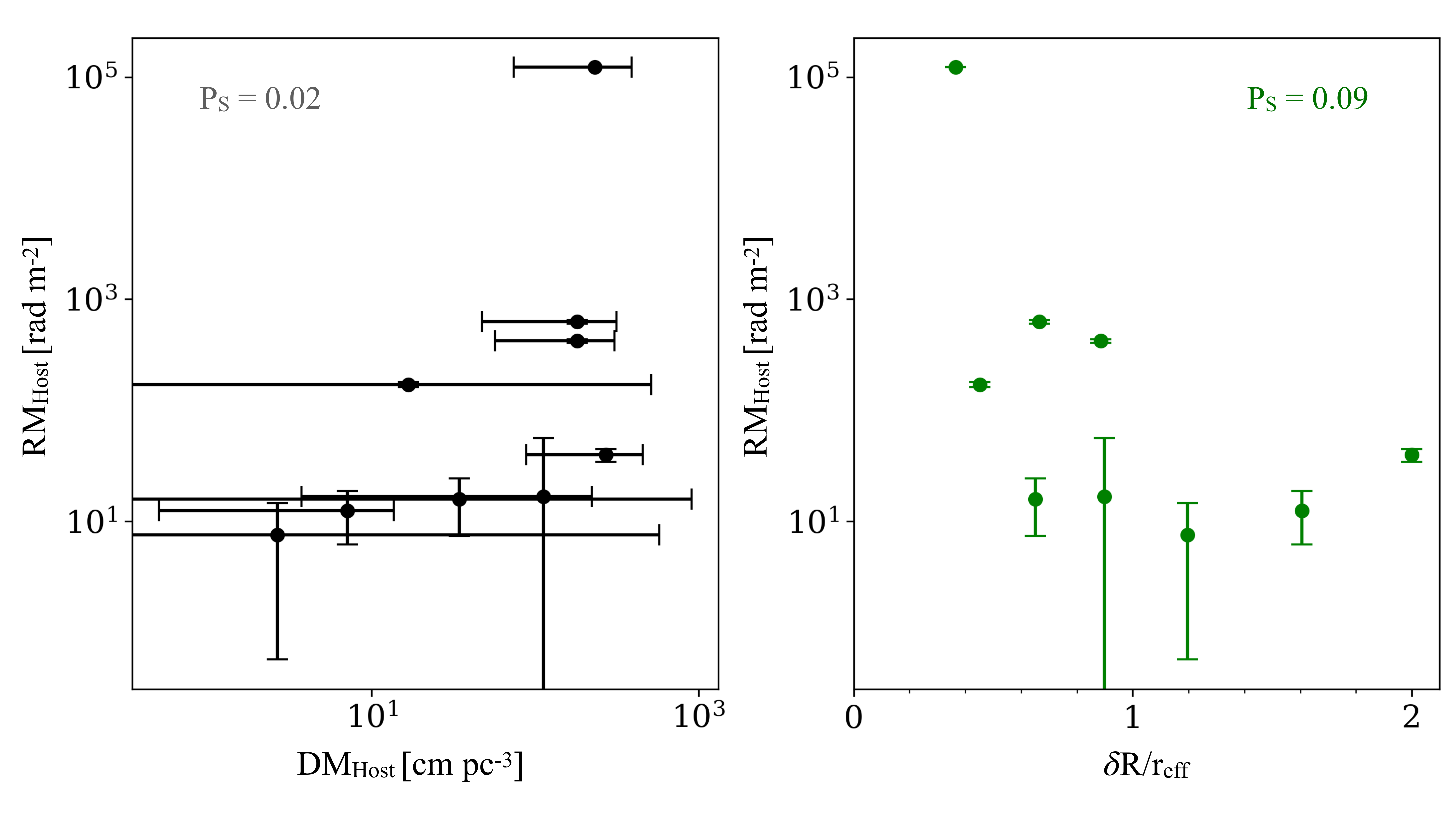}
    \caption{\rmhost as functions of \dmhost (\textit{left, black}) and \reloff (\textit{right, green}), with associated \pspear values. We can see a positive correlation by eye between \rmhost and \dmhost, however, a correlation between \rmhost and \reloff is not as apparent.
    %We have measurements (and simulations) that suggest that magnetic field strength has an inverse radial dependence; a claim which the anti-correlation between \reloff and \rmhost supports. This anti-correlation could also be due to a decrease in the density of ionized material as a function of galactic radius.
    }

    \label{fig:corr}
    \centering
\end{figure*}

\begin{figure*}[ht!]
    \centering
    \includegraphics[width=\linewidth]{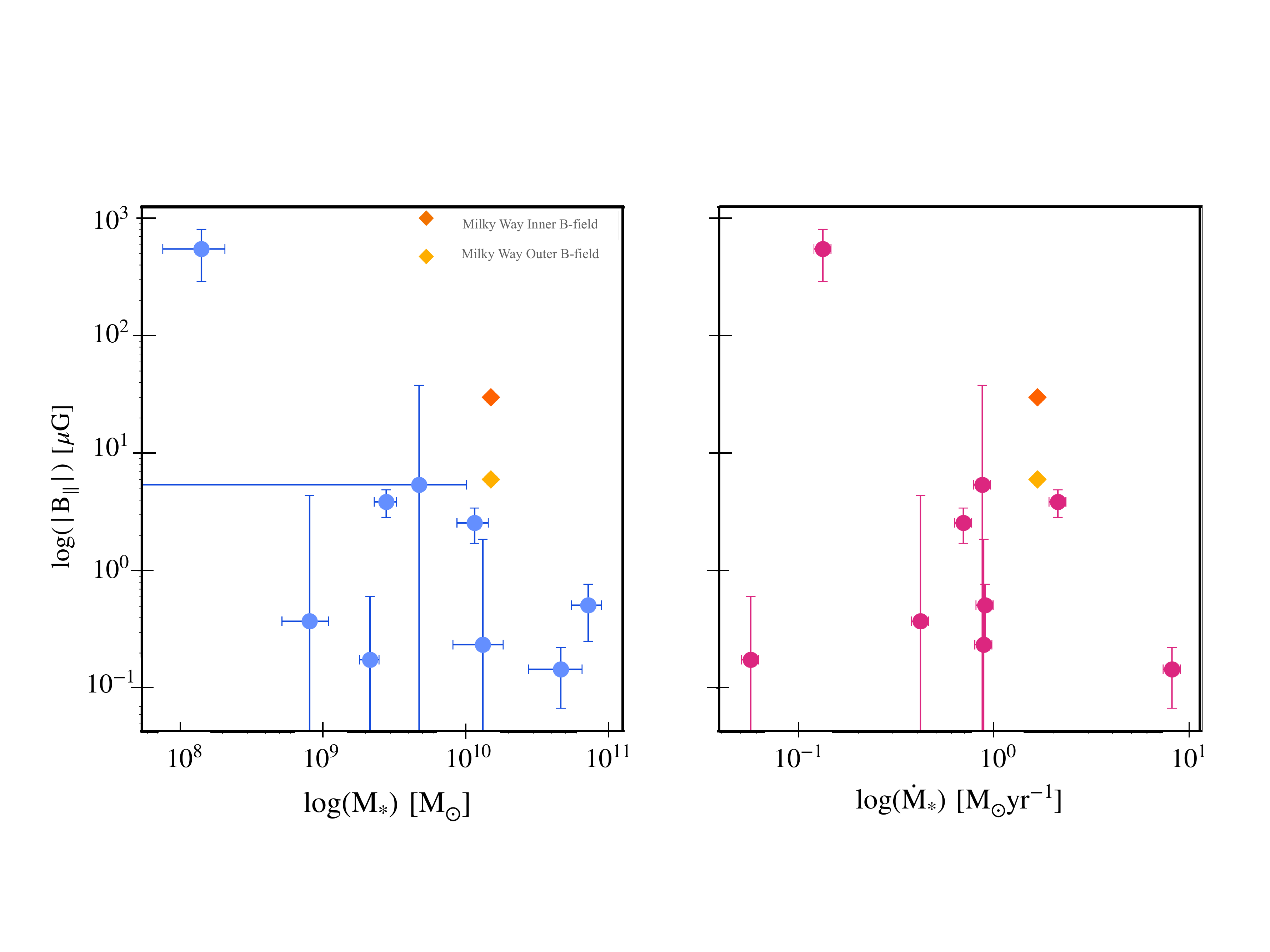}
    \caption{Comparison of the magnitude of the \bpar\ component of the magnetic field to stellar mass (right panel) and star-formation rates (left panel). We also show values determined for the Milky Way \citep{W&B2005}, for the bulge/inner field in \textit{yellow} and the disk/outer field in \textit{orange}, using values for mass and star-formation rate found in 
    \cite{Fragione_2017} and \cite{Licquia2015}, respectively. We find no apparent correlations between these global galactic characteristics and magnetic field measurements in this sample of FRB hosts.}
    \label{fig:b_mag}
\end{figure*}

The absolute values of the resultant \rspear\ values 
are shown in Figure~\ref{fig:radar}.
Although there are a number of parameters for which $|r_s| > 0.5$, the condition for significance is only satisfied 
for one quantity: \dmhost.
%We find the strongest correlations with \dmhost and \reloff, though in the case of \reloff, it is an anti-correlation. 
A positive correlation between \dmhost\ and \rmhost\ is in line with expectation because both quantities 
depend on the electron density of the host ISM. 
The correlation also lends further 
support to the assertion that \rmeg\ is 
dominated by \rmhost.

The next strongest correlation is an
anti-correlation between \rmhost\ and \reloff, 
although with $\pspear > 0.05$.
This trend follows 
observed and simulated inverse relationships between \bstrength\ and radius \citep[e.g.]{W&B2005, Pakmor2017}. %but we are unable to confirm with this current sample.
We await future observations to confirm (or refute)
such a trend in FRB observations. 
%We plot the two measures for which $|r_S| >0.5$ against \rmhost

As our sample is limited a sample size of \sampsize\ FRBs, such tests should be repeated with greater confidence with a larger sample size.
One also notes that the significance of these analyses 
is reduced by a trials-factor penalty incurred when 
testing for multiple correlations. 
%There are also a moderate but less significant correlations with \ssfr and \mstar. 

%Neither of these correlations meet our criteria of $P_S < 0.05$, but the significance of the correlation with \dmhost come close with $P_S = 0.06$.
%The significance of the \reloff correlation is moderate in comparison, which we believe is due to R1.
%In \textcolor{red}{Figure something}, we show \dmhost, \reloff, and \ssfr plotted against \rmhost, and we see a slight correlation by eye, though  Since the rotation measure for FRB\,20121102A is an outlier relative to the other values in this sample, the results of the correlation tests are slightly skewed. However, the correlations with \dmhost and \reloff still exist with moderate to high significance when R1 is disregarded.

%As our sample only contains \sampsize FRBs, such a test can be repeated with greater confidence with a larger sample size.
%Additionally, the significance is reduced by a trials-factor penalty incurred when testing for different correlations. 

% %%%%%%%%%%%%%%%%%%%%%%%%%%%%%%%%%%%%%%%%%%%%%%%%%%%%%%
\subsection{B-field Estimation}
\label{sec:Bfield}

As stated in \cite{Arshakian2009}, rotation measures of polarized background sources can be used to reconstruct a host or intervening galaxy's magnetic field topology. FRB signals have since been shown to be one such source. FRBs and their rotation measures can reveal the orientation and magnitude of ordered magnetic fields, making them a powerful tool. Here we look at the power of FRB sightlines to provide insight on the fields of the galaxies that host well-localized bursts.

As defined, \rotm\ is the sightline integral of
the parallel component of the magnetic field
weighted by the electron density.
Therefore the ratio of RM to DM (the unweighted
$n_e$ integral)
yields an estimate of the magnetic
field, after accounting for differences in the
prefactors \citep[e.g.][]{Akahori2016, Pandhi22}:

\begin{equation}
\begin{split}
    \langle B_{\parallel} \rangle &\approx \frac{\rm C_{D} RM}{\rm C_{R}DM} \\
    &= 12.3 \left ( \frac{\rm RM}{\rm 10 \, rad \, m^{-2} } \right )
    \left ( \frac{\rm DM}{\rm 10^{3} \, pc\, cm^{\rm -3}}
     \right )^{-1} \rm nG 
\end{split}
\label{eqn:b_par}
\end{equation}
 with $ \rm C_{\rm D}$ and $ \rm C_{\rm R}$ constants 
 equal to 1000 and 811.9, respectively. 
 In \cite{Akahori2016} they propose (with some 
improvements to equation \ref{eqn:b_par}) 
that many measurements of \rmfrb\ and 
\dmfrb\ can provide the data necessary to probe inter-galactic magnetic fields (IGMFs). This estimation also assumes a constant magnetic field and that there is no correlation between \edens and \bpar, whereas it is possible the magnetic field strength will likely decrease exponentially with radius and height in the disk (similar to $n_e$). 
These effects should be less than
%Though, the estimate is still good to 
an order of magnitude making the given ratio sufficient for estimating the 
magnitude of the magnetic fields in our sample's host galaxies. 
Furthermore, we adopt values of \rmhost\ and 
\dmhost\ to isolate the magnetic field estimation
to within the host galaxy.

Considering the turbulence and ionization due to active star-formation and the deaths of massive stars, one may suspect a relationship between star-formation rate and the magnitude of magnetic fields in the host. We also investigate the relationship between stellar mass and magnetic field strength. In \cite{Rodrigues19}---for galaxies at $z=0$---they find a slight positive correlation between galaxy mass and magnetic field, but note that this correlation is broken by a number of lower-mass hosts with much higher magnetic field strengths. 
In Figure~\ref{fig:b_mag}, we do not see a clear trend with strength \bstrength\ as a function of mass nor star-formation rate, in contrast to the predictions of \cite{Rodrigues19}. 
FRB\,20121102A exhibits the highest rotation measure of all the bursts, but resides in the lowest-mass host.
This is indicative of the highly magnetized environment that the burst progenitor lives within \citep{Michilli18}.

In most cases, the magnetic field strengths 
estimated are lower than values quoted for the Milky Way ($ \approx 6 \mu G$ in the outer reaches of the galaxy and $ 30 \mu G $ in the inner region towards the bulge). 
These Galactic values are dominated by the small-scale random fields which contrasts the regular fields along the line of sight that dominates RM. Comparing these two sets of measurements allows us to make distinctions between the strengths of various field components. We note that the values we have determined are largely consistent with $\simeq \mu G$ field magnitudes, which is broadly accepted as the general magnitude of large-scale regular fields measured in galactic disks \citep{Beck2019}.

% %%%%%%%%%%%%%%%%%%%%%%%%%%%%%%%%%%%%%%%%%%%%%%%%%%%%%%
\section{Modelling Galactic Magnetic Fields}
\label{sec:modeling}
\begin{figure}
    \centering
    \includegraphics[width=\linewidth]{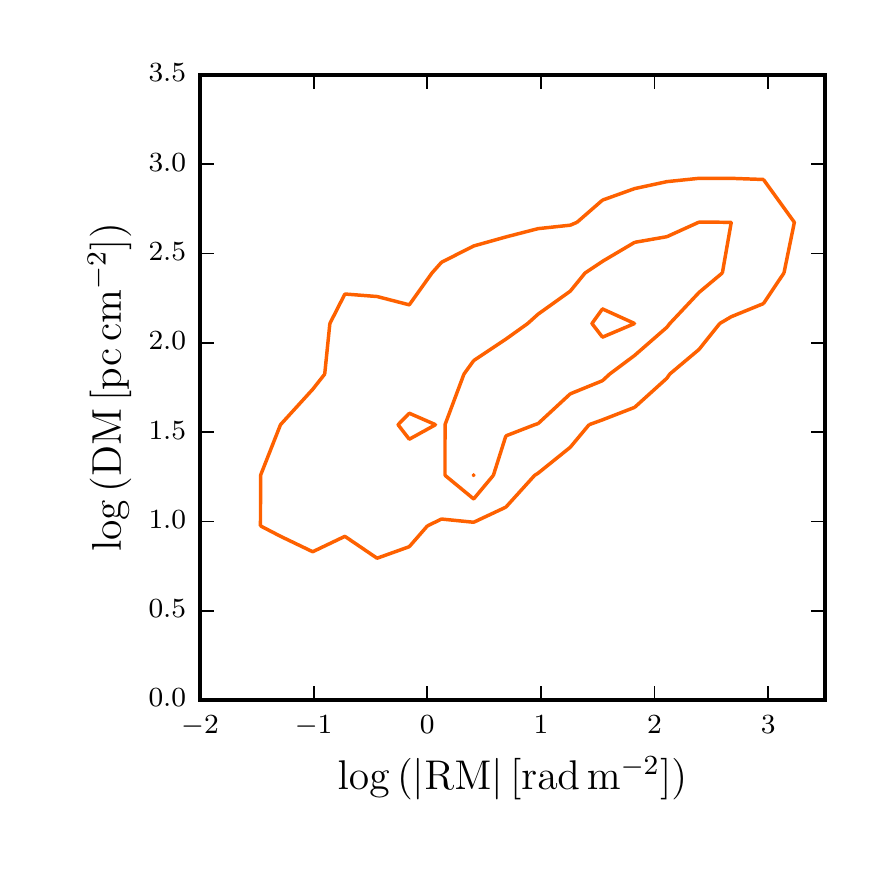}
    \caption{Correlation between RM and DM values for lines of sight from matched host galaxies in Auriga. We show $10\%$, $50\%$, and $90\%$ contours of the distribution. Although there is non-negligible scatter, the RM and DM are strongly correlated in the simulations.}
    \label{fig:aurigaRMvsDM}
\end{figure}

\begin{figure*}
    \centering
    \includegraphics[width=\linewidth]{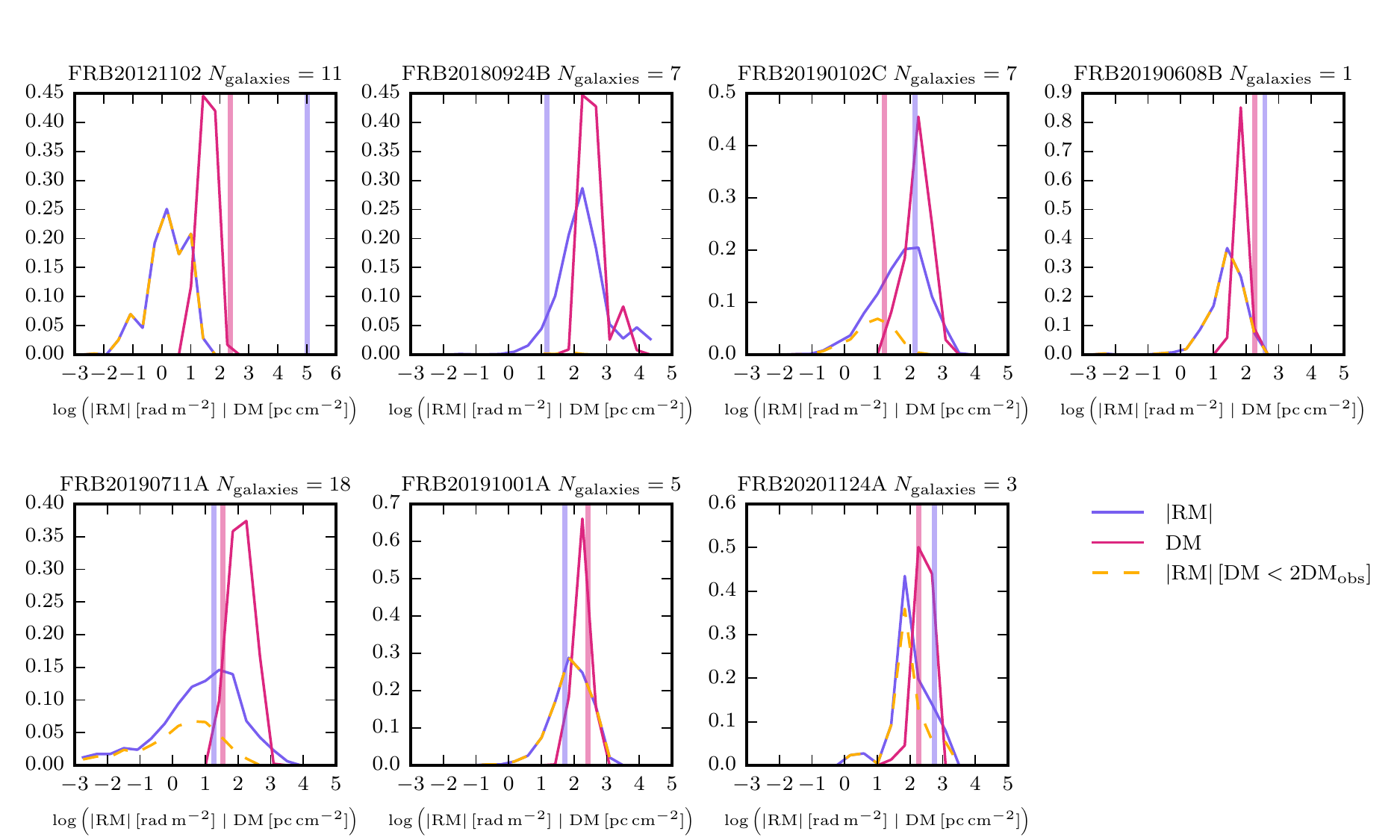}
    \caption{RM and DM distributions for lines of sight from matched host galaxies in Auriga. The name of the matched FRB host and the number of matching Auriga galaxies found are listed above each panel. Shown are histograms of the DM values (solid red lines) and absolute RM values (solid purple lines) along each line of sight. The vertical red and purple lines indicate the observed or derived \rmhost and \dmhostC values. We also show histograms of absolute RM values that include only lines of sight with a consistent DM value (dashed yellow lines; absent in panels where no consistent DM was found). The RM distributions predicted from the Auriga galaxies are generally in good agreement with the observed values, with the glaring exception of FRB20121102A whose RM value is far away from any lines of sight we find in Auriga. The modeled RM PDF for FRB20190608B is also noticeably lower than the observed value. The derived \dmhostC value for FRB20180924B is negative, therefore there is no vertical line shown here.}
    \label{fig:auriga}
\end{figure*}

\subsection {Auriga Model}
\label{sec:auriga}

To gain a better understanding of the physical mechanisms that can influence the magnetic fields in the host galaxies of our observed FRBs, we compare them to similar galaxies in the Auriga simulations.

The Auriga simulations are a set of cosmological magneto-hydrodynamical zoom-in simulations of Milky Way-like galaxies \citep{Grand2017}. They model magnetic fields in the approximation of ideal magnetohydrodynamics (MHD) using a second order finite volume scheme \citep{Pakmor2011,Pakmor2013} in the \textsc{Arepo} code \citep{Springel2010,Pakmor2016,Weinberger2020}. The magnetic field is initialised at $z=127$ as a uniform magnetic seed field with a strength of $1.6\times10^{-10}\, \rm physical~\mathrm{G}$. The simulation then evolves the magnetic field self-consistently until $z=0$.

When the galaxies first form, the magnetic field is quickly amplified via a turbulent small-scale dynamo and saturates before $z=4$ with a magnetic energy density that reaches $ \sim 10\%$ of the turbulent energy density and erases any information about the seed field in the galaxy \citep{Pakmor2014}. After the galaxies form a disk at $z<2$ the differential rotation in the disk leads to a second phase of magnetic field amplification that ends when the magnetic energy density reaches equipartition with the turbulent energy density \citep{Pakmor2017}. The magnetic field properties, synthetic Faraday rotation maps, and the magnetic field in the circum-galactic medium of the Auriga galaxies has been shown to be consistent with observations \citep{Pakmor2016,Pakmor2018,Pakmor2020}.

\begin{deluxetable*}{cccccccc}
\tablewidth{0pc}
\tablecaption{Estimated RM, DM and Magnetic Field of the Host Galaxies\label{tab:ci}}
\tabletypesize{\normalsize}
\tablehead{\colhead{FRB} & \colhead{RM Median}
& \colhead{DM Median} 
& \colhead{$\rm RM 95 \% \rm Interval$} 
& \colhead{$\rm DM 95 \% \rm Interval$} 
\\& (\rmunits)& (\dmunits)
 & (\rmunits) & (\dmunits)
 } 
\startdata 
20121102A$^\dag$& 2.02 & 2.02 & [0.04,18.56]& [0.04,18.56]\\ 
20180916B$^\dag$&&&&\\ 
20180924B& 146.22 & 57.31 & [6.01,12598.00]& [4.20,117.59]\\ 
20190102C& 53.55 & 7.24 & [0.63,1055.81]& [0.28,69.17]\\ 
20190608B& 25.84 & 25.84 & [1.80,120.41]& [1.80,120.41]\\ 
20190711A$^\dag$& 8.58 & 1.66 & [0.00,745.50]& [0.00,97.79]\\ 
20191001A& 77.71 & 77.71 & [3.40,635.87]& [3.40,635.87]\\ 
20200120E$^\dag$&&&&\\ 
20201124A$^\dag$& 86.13 & 67.51 & [2.83,837.41]& [2.04,789.91]\\ 
\hline 
\enddata 
\tablecomments{Daggers denote repeating FRBs. The data included here are taken from the Auriga simulations. No matches were found for FRBs\,20180916B and 20200120E, thus they have been left blank.} 
\end{deluxetable*}

The Auriga simulations focus on a Lagrangian high-resolution region with a typical radius of $1\,\mathrm{Mpc/h}$ around the central Milky Way-mass galaxies. This high-resolution region contains a large number of smaller galaxies without contamination from low-resolution dark matter particles that we also include in our sample here. We focus our analysis on the six high-resolution simulations of the Auriga project with a baryonic mass resolution of $\approx7-8\times10^{3}\,\mathrm M{_\odot}$. These are supplemented by yet unpublished simulations with the same mass resolution, centred on lower mass galaxies ($M_\mathrm{halo}=10^{10}-10^{11.5}$~M$_\odot$) for which the high-resolution regions also extend to about 5 times the virial radius around each central galaxy.

\subsection{Host and sightline selection}
\label{sec:auriga_select}

We first find galaxies in our simulation suite with stellar mass, star formation rate, and effective radius consistent with the host galaxies of our FRB sample. We calculate the stellar mass of a simulated galaxy by including all stars within three times its stellar half mass radius. Its star formation rate is averaged over the last 100~Myr using newly formed stars within the same radius. We use the stellar half mass radius as a proxy for the effective radius. We include all galaxies (both central and satellite galaxies) that match the FRB sample to within twice the observational error. For the effective radius, however, we add a 10 per cent error in quadrature to the observational error, because in some cases the derived errors were so small that no match could be found. With this selection procedure, we found one or more matching galaxies for $7$ out of the $9$ observed host galaxies.

We tilt each of the galaxies into the observed inclination and then integrate the RM and DM values for $256$ different lines of sight. The starting point of each line of sight is the position of a random star particle with an age younger than $200\,\mathrm{Myr}$. The frequent incidence of FRBs one or near spiral arms, could indicate the association of FRBs with relatively young stellar populations. From the starting point we integrate each line of sight until it reaches an observer at a distance of $100\,\mathrm{kpc}$. We checked that increasing the integration distance does not change our results.

The local electron density for the integration along the line of sight is computed exactly as described in \cite{Pakmor2018}, in particular assuming that only the volume-filling warm phase of the ISM contributes and that this phase is fully ionised. The magnetic field is taken directly from the simulation.

\subsection{Correlation between DM and RM}
\label{sec:rm_dm}

We show the correlation between RM and DM for all sightlines we computed from the Auriga galaxies in Figure~\ref{fig:aurigaRMvsDM}. RM and DM are clearly strongly correlated, as expected. 
However, there is significant scatter, i.e. for a fixed DM value the RM value can vary by two orders of magnitude. Nevertheless, the scatter is small on the scale of the overall variation of five orders of magnitudes in DM and eight orders of magnitude in RM.

Motivated by the strong correlation between DM and RM we not only compare the RM distributions of all lines of sight of matched FRB host galaxies, but also compare to a subsample of lines of sight that show consistent DM values.

\subsection{DM and RM of matched Auriga galaxies} \label{sec:rm_dm_auriga}
We show the distribution of RM (purple lines) and DM (red lines) values for $256$ lines of sights each of all galaxies in our sample consistent with the properties of the FRB host galaxies in Figure~\ref{fig:auriga}. We also show the measured values as vertical lines of the same color.

Strikingly, the shape of the distributions of synthetic DM and RM values match for most FRBs. The RM and DM distributions overlap with the values inferred for the FRB host galaxies from our observations with the exception of FRB20201124A. Median values and 95\% confidence intervals are shown in Table \ref{tab:ci}.

We also show the RM distributions restricted to lines of sights that have a DM value consistent with the observed host galaxy DM (${\rm DM}_{\rm sim} < max \left(100 \, \dmunits, 
2 \times \dmhostC \right)$, shown by the dashed, yellow curves). For most FRBs this restricted RM distribution is essentially the same as the full RM distribution. For FRB20190102C, FRB20201124A, and FRB20190711A this restriction reduces the high RM tail of the distribution. Interestingly, in all three cases the host RM estimated from observations lies on this tail that is reduced significantly by the restriction on DM. We also note (as seen in Figure~\ref{fig:auriga}) the noticeable difference between the modeled PDF and observed \rmhost for FRB20190608B. Though, again the value lands on the tail of the distribution. This could point to an non-negligible contribution of the local environment to the observed RM, as was discussed in \cite{Chittidi21}. The authors point out the high RM in comparison to other bursts such as FRB\,20180916, implying a magnetised local environment.

A larger sample is necessary to determine whether or not these variations are truly due to local effects.

An extreme exception is FRB20121102A. We do not find any lines of sight that have an RM value even remotely comparable to the large observed value. In contrast, the DM value is barely consistent with our synthetic lines of sight. This indicates that the magnetic field dominating the RM of FRB20121102A is part of its local stellar environment that is not included in our simulations. \cite{Michilli18} discusses this highly magnetized local environment. 

Note also, that it is likely increased scatter broadening of the FRB signal would bias against FRBs being detected with high host DM contribution.  

% %%%%%%%%%%%%%%%%%%%%%%%%%%%%%%%%%%%%%%%%%%%%%%%%%%%%%%%
\section{Discussion} \label{sec:discussion}

This sample represents the largest 
collection of FRB rotation measures presented with accompanying high-precision localizations and follow-up imaging of the associated hosts. 
A majority of the hosts in this sample are massive, star-forming galaxies at $z < 0.5$, with a few exceptions at lower mass or SFR.

To explore the relationships between FRB rotation measures and host characteristics, we first 
isolated the extragalactic contribution to the rotation measure (\rmeg).
We used the Galactic Faraday rotation map developed in \cite{Hutsch21} and found no correlation between \rmeg and $z$. 
This is consistent with measurements of the IGMF 
found in \cite{Carretti2022} which follows from
an expected
random nature and much lower strength of the fields in the IGM.  

We therefore disregarded IGMF contributions
and assert that \dmeg\ is dominated by the 
rotation measure of the host galaxy \dmhost.
We find a strong correlation between \rmhost\ and \dmhost, which supports this assertion and provides
encouraging confidence that FRBs probe the 
magnetic fields of their host galaxies.
This correlation is expected if the magnetic field has a significant ordered component that only varies weakly along the line of sight. Then both quantities depend similarly on the integrated density of the ionized medium along the line of sight through the host galaxy. This is consistent with our observational and theoretical picture of magnetic fields in massive disk galaxies \citep{Beck2015,Pakmor2017}.

There is evidence for an anti-correlation 
between host-normalized galacto-centric offset 
and \rmhost\ but at less than 95\%\ confidence.
A larger dataset is required to test whether
FRBs reveal this relationship, though
observed and modeled field strengths have been seen to show some radial dependence.

We considered several methods to isolate the host contribution to the dispersion measure \dmhost\ relating the emission measure to dispersion measure \citep{Reynolds1977} and applying the Macquart relation \citep{Macquart20} to estimate \dmhost. 
We then use the relation between dispersion and rotation measures described in papers such as \cite{Akahori2016} $\&$ \cite{Pandhi22} to make an estimate of \bpar\ for each of our host galaxies. 

With this method we find magnetic field strength estimates for our sample of the order of $\sim 1 \mu G$. The estimate, however, disregards field reversals of the magnetic field along the line of sight as well as differences in the exponential scaling of the magnetic field strength and electron densities with radius and height in the disk. Therefore, although the values are lower than values quoted for the Milky Way (see Figure \ref{fig:b_mag}), they are better seen as lower limits and are fully consistent with general expectations for galactic magnetic field strengths \citep{Beck2015}.
The uncertainty in our determinations of \dmhostC could have some effect on the derived \bpar, but would not cause a notable increase.

Four (possibly 5) of the hosts in this sample exhibit spiral structure, and the bursts originate on or near the spiral arms. According to \cite{Beck2015}, 
the strongest ordered fields are found in inter-arm regions due to shear caused by differential rotation and a large scale dynamo that operates preferentially in the inter-arm regions. Because of the preferred location of
FRBs on/near spiral arms, it is possible that our field strengths are referring to medium-scale ($\sim 1$\,kpc) regular fields that are affected by turbulence in the spiral arms. 
 Figure~\ref{fig:bstrength} shows where our measurements lie with respect to lines of constant \bpar\ of varying magnitudes 
 (with the values we derived shown in Table \ref{tab:B}). These values align well with the average strengths of large-scale regular fields 
 \citep[on scales of 5-10 kpc][]{Beck2019}, 
 where the large scale rotation sets the strength and structure of the magnetic field.

\begin{figure}
    \centering
    \includegraphics[width=\linewidth]{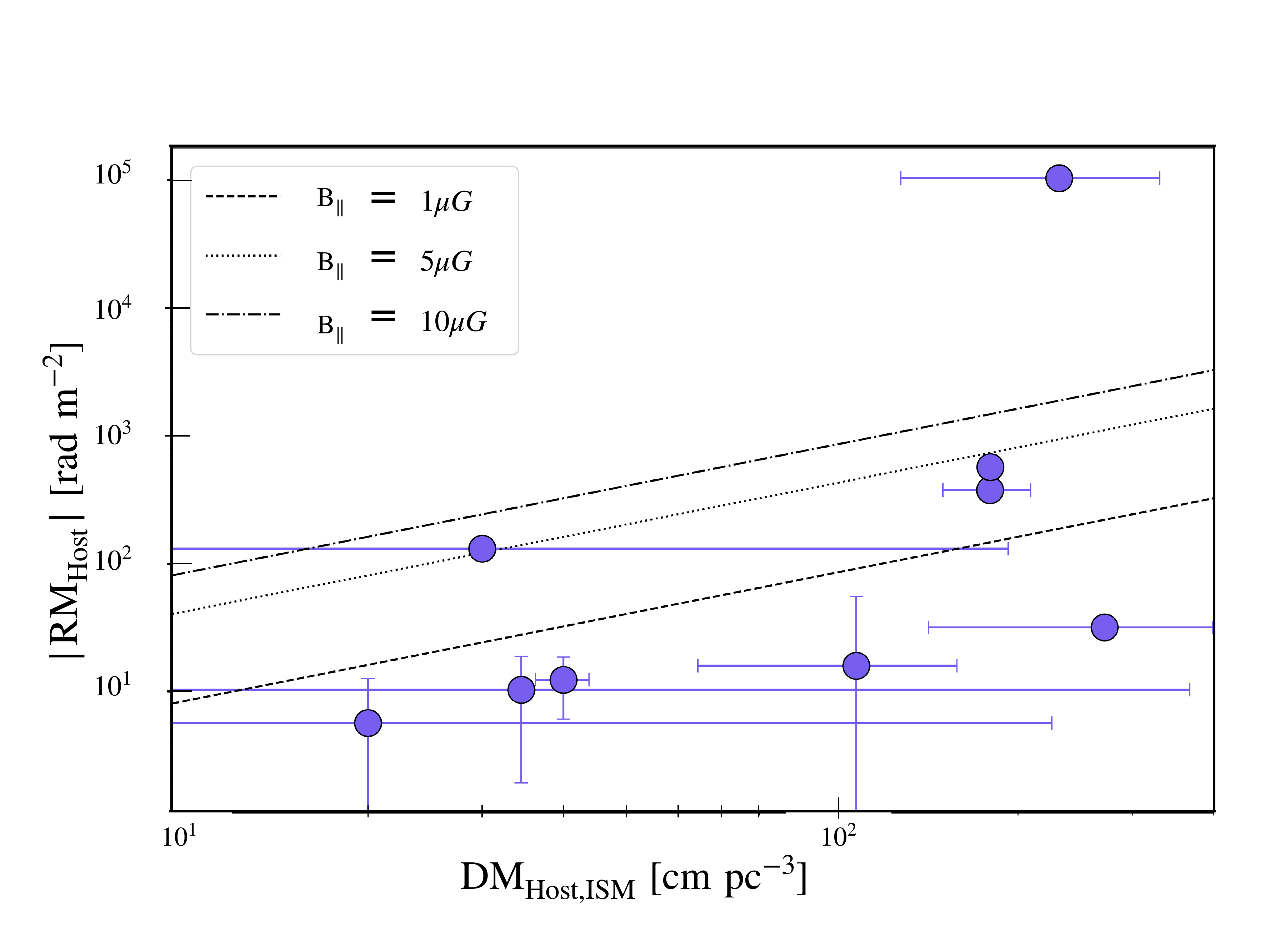}
    \caption{\rmhost and \dmhostI with lines of constant magnetic field strength (1 $\mu G$, dashed; 5 $\mu G$, dotted; 10 $\mu G$, dash-dotted). We calculate the value of \bpar for each of the FRBs using Equation \ref{eqn:b_par}. The majority of the values are less than 5 $\mu \rm G$, with many even falling below 1 $\mu \rm G$. There is one outlier, with a field magnitude far exceeding 10 $\mu \rm G$. These values are relatively consistent with $\mu \rm G$ fields, but are at the lower end of the range of 1-15 $\mu G$.}
    \label{fig:bstrength}
\end{figure}

Using forward modelling of cosmological simulations instead, we also find that observed \rmhost\ and therefore \bpar, are consistent with the Auriga simulations (see Figure~\ref{fig:auriga}). There is the notable exception of FRB\,20121102A, which we know is embedded in a highly magnetized environment. The predicted RMs for this FRB were not able to approach the observed value, as any contribution from local environments was
not included. 
This provides some hope that with a sufficient number of FRBs with polarization data and $\sim mas$ localizations, we will be able to disentangle the ISM and local environment contributions to the RM, and provide constraints on each. 

Limiting the simulated sightline-selection to those more consistent with the observed DM for each burst reduces the tail towards high RM values of the predicted RM distribution for three of the galaxies. Although the distribution is still consistent with the observed values, the majority of predicted RM values fall below the observed ones. This could imply a non-negligible contribution from the local stellar environment of the FRB. 

Combining FRB RM signals with measures of synchrotron polarization and estimates using galactic Zeeman
effect measurements --- which characterize the ISM magnetic field --- may also help to disentangle the magnetic field contributions within the host galaxy.

Finally, we find insignificant correlations with extant properties such as \reff, SFR, \msd, \mstar, shown in Figure \ref{fig:radar}. With larger, upcoming surveys, these relationships can be explored in more detail with higher statistical power. There is also no apparent relationship between FRB repetition and the host and environmental characteristics we have explored in this paper. There also seems to be little differentiation in the sample presented in \cite{Gordon23} where they explore the overall characteristics and star-formation histories of FRB hosts.
 This is in contrast to papers such as \cite{Pleunis20} which point out that there are some marked differences in burst characteristics (such as bandwidth and duration) of repeating and non-repeating FRBs. %Increased sample sizes are a must to truly disentangle how burst differences map to differences in host characteristics. 

We plan to repeat and expand
this study with a larger sample of more precisely localized bursts with accompanying high-resolution imaging and spectroscopy. More data would not only aid in the narrowing of possible progenitors, we can also learn about galactic magnetism and its effects on galaxy formation and evolution. With the onset of large-scale surveys such as CRAFT with the upcoming CRACO upgrade, the number of FRBs that meet these criteria will vastly increase ($\sim 3$ FRBs per day!), and help to determine what, on average, the local environments of FRBs look like in terms of stellar populations, magnetism and more, and investigate (inter-)galactic magnetism over cosmological time!

\section*{Acknowledgements}
The authors would like to thank the referee and Ranier Beck for their helpful comments on this work. A.G.M. acknowledges support by the National Science Foundation Graduate Research Fellowship under Grant No. 1842400. Authors A.G.M., J.X.P., S.S, M.R.,
and N.T., as members of the Fast and Fortunate for FRB
Follow-up team, acknowledge support from 
NSF grants AST-1911140, AST-1910471
and AST-2206490.
This work is supported by the Nantucket Maria Mitchell Association. 
This research was supported in part by the National Science Foundation under Grant No. NSF PHY-1748958.
N.T. acknowledges support by FONDECYT grant 11191217. 
R.M.S. acknowledges support through Australian Research Council Future Fellowship FT190100155.
F.v.d.V. is supported by a Royal Society University Research Fellowship (URF\textbackslash R1\textbackslash 191703).
This research is based on observations made with the NASA/ESA Hubble Space Telescope obtained from the Space Telescope Science Institute, which is operated by the Association of Universities for Research in Astronomy, Inc., under NASA contract NAS 5–26555. These observations are associated with programs 15878, 16080, and 14890. Support for Program numbers 15878 and 16080 were provided through a grant from the STScI under NASA contract NAS5- 26555.

\bibliography{hst_refs}

\end{document}